\documentclass[11pt,reqno]{amsart}      

\usepackage{amssymb}
\usepackage{amsthm}
\usepackage{amsmath}

\usepackage[dvips]{graphicx}
\usepackage{psfrag}
\usepackage{subfigure}

\usepackage{a4}

\numberwithin{equation}{section}        

\newcommand{\Ric}{\text{\rm Ric}}

\newcommand{\XX}{\mathbf X }

\newcommand{\UU}{\mathbf{U}}
\renewcommand{\SS}{\mathbf{S}}
\renewcommand{\AA}{\mathbf{A}}
\newcommand{\textfrac}[2]{{\textstyle \frac{#1}{#2}}}

\newcommand{\FA}{{\text{(F$_\text{A}$)}}}
\newcommand{\FB}{{\text{(F$_\text{B}$)}}}
\newcommand{\Fstar}{{\text{(F$_*$)}}}

\theoremstyle{plain}
\newtheorem{thm}{Theorem}[section]

\newtheorem{cor}[thm]{Corollary}

\newtheorem{prop}[thm]{Proposition}

\theoremstyle{remark}
\newtheorem*{remark}{Remark}
\newtheorem*{coro}{Corollary}

\allowdisplaybreaks[2]

\begin{document}

\title{Eternal acceleration from M-theory}

\author[L.~Andersson]{Lars Andersson}

\address{Max Planck Institute for Gravitational Physics \\
Albert Einstein Institute\\
Am M\"uhlenberg 1\\
D-14476 Golm\\
Germany \\
\and 
Department of Mathematics\\
University of Miami\\
Coral Gables, FL 33124\\
USA}

\email{larsa@math.miami.edu}

\thanks{Supported in part by the NSF, contract no.\ DMS 0407732.}

\author[J.M.~Heinzle]{J.\ Mark Heinzle}

\address{Max Planck Institute for Gravitational Physics \\
Albert Einstein Institute\\
Am M\"uhlenberg 1\\
D-14476 Golm\\
Germany}

\curraddr{Institute for Theoretical Physics \\
University of Vienna \\
Boltzmanngasse 5\\
A-1090 Vienna \\
Austria}

\email{mark.heinzle@aei.mpg.de}





\begin{abstract}
The dimensional reduction of $D$-dimensional spacetimes arising in
string/M-theory, to the conformal Einstein frame, may give rise to
cosmologies with accelerated expansion. 
Through a complete analysis of the
dynamics of doubly warped product spacetimes, in terms of 
scale invariant variables, it is demonstrated  
that for $D \geq 10$, eternally accelerating 
4-dimensional $\kappa = -1$ Friedmann cosmologies arise from
dimensional reduction on an internal space with negative Einstein geometry. 
\end{abstract}
\maketitle


\section{Introduction}

The current standard model of cosmology has as an essential element the
accelerated expansion of the universe.
In order to achieve accelerated expansion 
the strong 
energy condition must be violated. 
Numerous matter models which provide accelerated expansion 
have been proposed, among them
the cosmological constant $\Lambda$ and scalar field models like quintessence
and $k$-essence. 

It has been shown by
Townsend and Wohlfarth \cite{Townsend/Wohlfarth:2003} by considering
a Kaluza-Klein reduction of a $D$-dimensional spacetime with a 
hyperbolic internal space that 
the dimensionally reduced universe 
may exhibit a period of accelerated expansion, 
even though the $D$-dimensional model one starts from does
not violate the strong energy condition.
This result thus circumvents 
a 'no-go' theorem which states that
dimensional reduction of supergravity models arising from string/M-theory (which in particular satisfy the
strong energy condition) cannot give rise to cosmologies with 
accelerated expansion in case the internal
geometry is time-independent \cite{Gibbons:1985,Maldacena/Nunez:2001}.

The $D$-dimensional models discussed in~\cite{Townsend/Wohlfarth:2003} are 
vacuum spacetimes which are 
warped products of a $(3+1)$-dimensional $\kappa = 0$ Friedmann
model with an 
$n$-dimensional hyperbolic space, and are thus examples of Lorentzian 
doubly warped products.
The dimensionally reduced model exhibit a transient phase of acceleration.
In~\cite{Chen/Ho/Neupane/Ohta/Wang:2003} it has been shown that 
it is possible to obtain late time accelerated expansion,
if the $(3+1)$-dimensional model is a $\kappa = -1$ Friedmann model.
In the same paper it was 
also argued, based on a perturbation argument,
that solutions with eternal acceleration might exist.

In this paper we apply the method of scale invariant dynamics
to the study
of vacuum doubly warped products and their dimensional reduction. In particular, we
prove the existence of 
eternally accelerating solutions for $D \geq 10$.
Our analysis shows
that eternally accelerating cosmologies cannot be obtained from 
perturbations of the (dimensionally reduced) $D$-dimensional Friedmann model,
as was argued in~\cite{Chen/Ho/Neupane/Ohta/Wang:2003}.
For $D < 10$ we prove that no eternally accelerating models can occur. 

We consider line elements on a $D = m+n+1$ dimensional spacetime  
$\mathbb{R} \times M \times N$ of the form 
\begin{equation}\label{eq:Dgeom} 
-dt^2 + a^2(t) g + b^2(t) h\:,
\end{equation} 
where $(M,g)$, $(N,h)$, are $m$- and $n$-dimensional Einstein spaces
of non-negative Einstein curvature. (Positive curvature is discussed in Sec.~\ref{mpsec}.)
The vacuum Einstein equations give a
system of ODEs for $a$, $b$; 
however, in our approach, we formulate the equations in scale
invariant variables and obtain a regular dynamical system on a compact state space, cf.~Sec.~\ref{sec:prelim}. 
The state space is topologically a disk; in the interior
there are represented models where both factors $g$, $h$ have negative
Einstein curvature; solutions on the boundaries
correspond to models where one of the factors is Ricci flat.

For the $D$-dimensional geometry, we
show that the behavior is asymptotically Friedmann in the expanding
direction, 
corresponding to the approach to a stable fixed
point (F$_*$) in the formulation of the scale invariant dynamics,  
and asymptotically Kasner-like in the collapsing direction. The
asymptotic behavior in the expanding direction depends on the dimension; for
$D < 10$, the Friedmann point $\Fstar$ is a stable spiral point, while for $D
\geq 10$, $\Fstar$ is a stable node. The system exhibits two qualitatively
different types of orbits for the collapsing direction: the generic orbit is
asymptotically Kasner like with non-vanishing generalized Kasner exponents,
which implies that in the direction of the singularity 
one of the factors in the doubly warped
product is expanding while the other is collapsing. This case is analogous
to the vacuum Kasner spacetime with line element $-dt^2 + t^{2p} dx^2 +
t^{2q} dy^2 + t^{2r} dz^2$ and with exponents $(p,q,r) = (2/3, 2/3, -1/3)$. 
In addition, there are
two exceptional orbits, asymptotic to one of two fixed points 
$\FA,\FB$ on the boundary of the state space, 
with the property that one of the factors converges to a constant
and the other scales asymptotically with proper time toward the
singularity. 
These cases are 
analogous to the flat Kasner spacetime with exponents $(1,0,0)$. The
analysis leading to this description 
is performed in Section~\ref{sec:analysis}, in particular,
Theorem~\ref{mm1thm} summarizes the main results. 

The issue of whether the $D$-dimensional models considered in this paper
are stable in the sense that they admit large families of perturbations with
quiescent behavior at the singularity is discussed in Section~\ref{sec:AVTD}. 
We find that for $D \geq 11$, this is indeed the case. This
agrees with the general analysis of 
\cite{demaret:etal:1985, damour:etal:kasnerlike}. Thus we see an
indication in this simple context that spacetimes with $D=11$ exhibits
special features. 

In Section~\ref{acceleration} we apply the results about 
the $D$-dimensional scale invariant dynamics 
to the dynamics of the dimensionally reduced models. We
take the factor $(N,h)$ as internal space. It
turns out that a cosmological model that arises through dimensional
reduction exhibits accelerated expansion when the corresponding solution of the
scale invariant dynamical system lies 
in a certain region $\AA$ of the state space, cf.~Fig.~\ref{mm1accelall}. 
This domain of acceleration $\AA$ 
intersects the boundary of the state space in an open interval; hence,
solutions on the boundary (which describe models where $g$ is Ricci flat)
exhibit a phase of acceleration, which explains the transient acceleration 
found by Townsend and Wohlfarth~\cite{Townsend/Wohlfarth:2003}.
Furthermore we find that 
the Friedmann fixed point $\Fstar$ lies on the 
boundary of $\AA$. 
The fact that, for $D < 10$, the Friedmann
point $\Fstar$ is a spiral point thus entails that
models cannot exhibit
late time accelerated expansion, even with negatively curved spatial slices
and hyperbolic internal space. Instead one finds that cosmologies which arise
through dimensional reduction of spacetimes with  $D < 10$ have an infinite
sequence of episodes with accelerating and decelerating expansion. 
However, for $D \geq 10$, based on arguments combining the local properties
of $\Fstar$ and global properties of scale invariant dynamical system,
we are able prove that there exists a unique dimensionally reduced cosmology with eternal
acceleration (i.e., a cosmology that expands at a accelerated rate for all times),
cf.~Theorem~\ref{eternalthm}.
Note, however, that uniqueness of this model amounts to the model being unstable
under any perturbation; the behavior of generic models is described in Corollary~\ref{fullcor}.

Finally, in Section~\ref{mpsec} we consider the case where one of the factor in~\eqref{eq:Dgeom}
has positive Einstein curvature and one negative curvature.
Using again a formulation in terms of scale invariant variables, we
are able to perform a complete analysis of the dynamics of the
$D$-dimensional geometries as well as their dimensional reduction. 
The domain of acceleration
can be found explicitly, and it turns out that there exist models with one or
two epochs of expansion (whereof one can be partly accelerating), 
but that all models undergo recollapse to a
big crunch.

\section{Preliminaries} \label{sec:prelim} 

Let $(M,g)$ and $(N,h)$ be Einstein manifolds of dimension $m$ and $n$ with 
\begin{equation}
\Ric_g = k_g (m+n-1) g\:,\qquad\qquad
\Ric_h = k_h (m+n-1) h\:;
\end{equation}
$k_g$ and $k_h$ are constants taking values in $\{+1,0,-1\}$. 
On the $D=1+m+n$ dimensional spacetime $\mathbb{R} \times M \times N$ consider a line element of the form of a doubly warped product
\begin{equation}\label{metric}
-dt^2 + a^2(t) \,g + b^2(t)\, h\:,
\end{equation}
where $a>0$ and $b>0$ without loss of generality,
and impose the vacuum Einstein equations. Let $i,j$ be indices running over
$1\dots m+n$. The second fundamental form is 
\[
K_i^{\ j} = \left\{ 
\begin{array}{ccc} -(\dot{a}/a)\: \delta_i^{\ j} & \quad &1\leq i,j\leq m\\ 
-(\dot{b}/b)\:  \delta_i^{\ j} & \quad & m+1 \leq i,j \leq m+n\:,
\end{array} \right.
\]
where we use the notation $\dot f = \partial_t f$. 
Let $p = -\dot a/a$, $q = -\dot b/b$. Then
the mean curvature is $H = \mathrm{tr} K = m p+n q$, and the Einstein evolution equations imply 
\begin{subequations}
\begin{align} 
\dot p & = p H +k_g \frac{m+n-1}{a^2} \\
\dot q & = q H +k_h \frac{m+n-1}{b^2} \:.
\end{align} 
\end{subequations}
From these evolution equations we obtain
$$
\dot H = m p^2 + n q^2 \:.
$$
Here we have employed the Hamiltonian constraint equation, 
$$
0 = k_g \frac{m(m+n-1)}{a^2} +k_h \frac{n(m+n-1)}{b^2} + H^2 - (m p^2 + n q^2) \:.
$$
Note that $k_g \leq 0$ and $k_h \leq 0$ entails $H \neq 0$; in the following we focus on this case, 
the case $k_g \geq 0$, $k_h \leq 0$ is treated in Sec.~\ref{mpsec}.
We now introduce scale invariant variables $(P,Q,A,B)$ according to
\begin{equation}\label{vartrans}
P = \frac{p}{H} \:,\qquad Q= \frac{q}{H} \:, \qquad
A = -\frac{1}{a H}\:, \qquad B = -\frac{1}{b H} \:.
\end{equation}
The variables $A,B$ are curvature quantities.
By definition we obtain $m P + n Q =1$, and the Hamiltonian constraint now reads 
$$
1 = -(m+n-1)(k_g m A^2 + k_h n B^2) + (mP^2 + nQ^2) \:.
$$
In the following we introduce the time $\tau$ by 
$\partial_\tau = H^{-1} \partial_t$; we will use a prime ${}^\prime$ to denote differentiation w.r.t.\ $\tau$.
Note that with our conventions $H < 0$, hence by introducing the time $\tau$ 
we have the singularity to the future.  \\

\section{Analysis} \label{sec:analysis} 

The variable transformation $(a,b,p,q) \mapsto (A,B,P,Q,H)$ enables us to write
the Einstein equations 
as a system of 
evolution equations, 
\begin{equation}\label{eq:H-evol}
H' = H (m P^2 + n Q^2) 
\end{equation} 
and 
\begin{subequations} \label{eq:evol} 
\begin{align} 
\label{eq:A} A^\prime &= A \,[P - (m P^2 + n Q^2)]  \\
\label{eq:B} B^\prime &= B \,[Q - (m P^2 + n Q^2)]  \\
\label{eq:P} P^\prime &= P \,[1 - (m P^2 + n Q^2)]  + (m+n-1) k_g A^2 \\
\label{eq:Q} Q^\prime &= Q \,[1 - (m P^2 + n Q^2)]  + (m+n-1) k_h B^2 \:,
\end{align} 
\end{subequations} 
supplemented by two constraint equations, 
\begin{subequations} \label{eq:con} 
\begin{align} 
\label{eq:trace} 
C_1 & = m P + n Q - 1 = 0 \\
\label{eq:Ham}
C_2 & = (m P^2 + n Q^2) - (m+n-1)(k_g m A^2 + k_h n B^2) -1 =0 \:.
\end{align} 
\end{subequations} 
Since Eq.~\eqref{eq:H-evol} decouples, the entire dynamics is encoded in the
reduced dynamical system~\eqref{eq:evol}, which is autonomous and 
regular for all $(A,B,P,Q) \in \mathbb{R}^4$.
From~\eqref{eq:A} and~\eqref{eq:B} it follows that $A>0$ and $B>0$ are
invariant under the flow of the system; henceforth, without loss of generality, 
we will always impose these conditions, i.e., we consider
the dynamical system~\eqref{eq:evol} on the state space 
$\mathbf{X} = \{ (A,B,P,Q)\:|\: (A>0)\wedge (B>0)\}$.

In our subsequent analysis it will turn out that there exist solutions
with $A B \rightarrow 0$ for $|\tau| \rightarrow \infty$.
This suggests to include the boundaries $A=0$ and $B=0$ of the 
state space in our dynamical systems analysis; note that 
the system~\eqref{eq:evol} can be smoothly extended to $\bar{\XX}$.

The equations on the invariant subset $A=0$ (respectively $B=0$) can be interpreted
as the reduced system of coupled equations that arises when the first factor
(respectively the second factor) of the metric~\eqref{metric} is Ricci flat.
This is because
setting $A=0$ in~\eqref{eq:evol} and~\eqref{eq:con}
corresponds to setting $k_g =0$ and discarding the decoupled equation for $A$.
(In the case $k_g =0$, due to the decoupling, the equation for $A$ does not carry any
dynamical information; note, however, that 
the equation must be added in order to reconstruct the original variables.)

By construction the constraints~\eqref{eq:con} are preserved during the evolution.
To see the propagation of constraints explicitly we compute
\begin{subequations}
\begin{align} 
C_1^\prime & = [ 1 -(m P^2 +n Q^2) ]\: C_1 - C_2 \\
C_2^\prime & = -2 (m P^2 + n Q^2)\, C_2 \:.
\end{align} 
\end{subequations} 
Hence the physical state space 
\begin{equation}
\SS := \big \{ (A,B,P,Q)\:|\: (C_1 = 0)\wedge (C_2 = 0) \wedge  (A>0) \wedge (B>0) \big\}
\end{equation}
and the closure $\bar{\SS}$ are invariant subsets of $\XX$ (and $\bar{\XX}$, respectively).

\begin{figure}[Ht]
\begin{center}
\psfrag{P}[cc][cc][1][0]{$P$}
\psfrag{A}[cc][cc][1][0]{$A$}
\psfrag{B}[cc][cc][1][0]{$B$}
\psfrag{A0}[cc][cc][0.8][-65]{$A=0$}
\psfrag{B0}[cc][cc][0.8][65]{$B=0$}
\psfrag{Pc}[lc][cc][0.8][0]{$P=\frac{1}{m+n}$}
\includegraphics[width=0.5\textwidth]{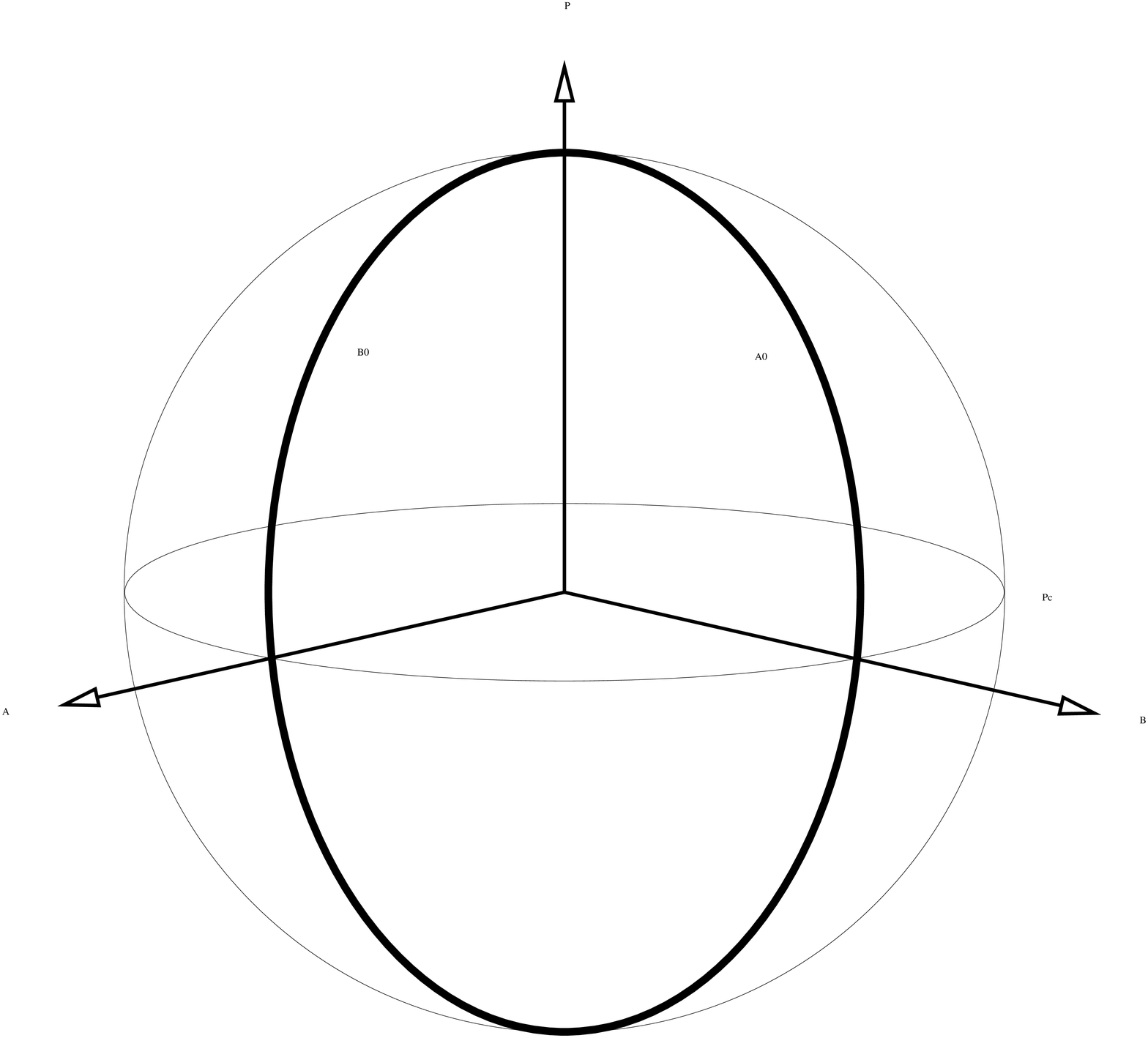}
\caption{The state space $\bar{\SS}$.}
\label{statespace}
\end{center}
\end{figure}

When the variable constraint~\eqref{eq:trace} is solved for $Q$, the Hamiltonian constraint~\eqref{eq:Ham} becomes 
\begin{equation}\label{ellipsoid}
\frac{m}{n} (m+n) \Big( P - \frac{1}{m+n}\Big)^2 - (m+n-1) (k_g m A^2 + k_h n B^2) = \frac{m+n-1}{m+n}\:.
\end{equation}
When $k_g = k_h =-1$, this condition defines
an ellipsoid centered at $(A,B,P) = (0,0,1/(m+n))$.
Hence, 
$\bar{\SS}$
corresponds to a quarter ellipsoid, which is topologically a disk; see Fig.~\ref{statespace}.
In particular, we see that in the case $k_g = k_h = -1$ the state space $\SS$ has a compact closure.
(Recall that the dynamical system induced on the boundaries of $\SS$ represents the reduced dynamics 
of the cases $k_g =0$, $k_h=-1$ and $k_g =-1$, $k_h=0$.)

In the subsequent sections, \boldmath $k_g = k_h = -1$ \unboldmath is understood
(the case $k_g k_h = -1$ is treated in Sec.~\ref{mpsec}). 
This entails that the dimensions of the factors $M$ and $N$ satisfy 
\boldmath $m>1$ and $n>1$. \unboldmath 

\subsection{Equilibrium points and invariant subsets}

An elementary analysis shows that 
the equilibrium points in $\bar{\SS}$ of the autonomous system~\eqref{eq:evol} are 
\begin{subequations}
\begin{align} 
\label{F12}
(\text{F}_{1,2})&: \quad A = B = 0, \quad m P^2 + n Q^2 = 1, \quad m P+n Q=1\:, \\
(\text{F}_{\mathrm{A}})&: \quad A = 0, \quad B^2 = \frac{n-1}{n^2} \frac{1}{m+n-1}, \quad P =
0,\quad Q = \frac{1}{n} \, , \\
\label{FB}
(\text{F}_{\mathrm{B}})&: \quad A^2 = \frac{m-1}{m^2} \frac{1}{m+n-1}, \quad B = 0 , \quad P =
\frac{1}{m}, \quad Q = 0 \, ,\\ 
(\text{F}_*)&: \quad A =  B = P = Q = \frac{1}{m+n} \, .
\end{align} 
\end{subequations}
Note that in~\eqref{F12} the equations $m P^2 + n Q^2 = 1$ and $m P+n Q=1$ possess two solutions $(P_1, Q_1)$ and $(P_2, Q_2)$,
which define the fixed points $\text{F}_1$ and $\text{F}_2$, respectively; we have $P_1 > 0$ ($Q_1 < 0$) and $P_2 < 0$ ($Q_2 > 0$); 
see Fig.~\ref{mm1fig}.
Note that $|P_i| < 1$ and $|Q_i|< 1$, because $m>1$ and $n>1$. 

As noted above, the boundaries of the state space are invariant subsets.
We denote $\UU_{\mathrm{A}} = \{A=0\} \cap \bar{\SS}$ and
$\UU_{\mathrm{B}} = \{B=0\} \cap \bar{\SS}$.
The set $\UU_{\mathrm{A}}$ contains the fixed points $\text{F}_1$, $\text{F}_2$, and $\text{F}_A$;
$\UU_{\mathrm{B}}$ contains  $\text{F}_1$, $\text{F}_2$, and $\text{F}_B$.
Recall that these invariant subsets describe the dynamics of a spacetime 
with spatial geometry which is a product of an Einstein
manifold with a Ricci flat space.

\subsection{Stability of the fixed points}

\quad\newline
\noindent\textbf{The case $(\text{F}_{1,2})$}:
Since the fixed points $(\text{F}_{1,2})$ on $\bar{\SS}$ are given as intersections of the
one-dimensional invariant subspaces $\UU_{\mathrm{A}}$ and $\UU_{\mathrm{B}}$,
the eigenvectors of the linearization of the system~\eqref{eq:evol}
must be tangential to $\UU_{\mathrm{A}}$ and $\UU_{\mathrm{B}}$.
For $(\text{F}_{i\,})$, the associated eigenvalues are $(P_{i}-1)$ and $(Q_{i}-1)$, which follows
from $A^{-1} A^\prime |_{(\text{F}_{i})} = P_{i} -1$ and the analogous relation
for $B^{-1} B^\prime$.
Since $|P_{i}|<1$ and $|Q_{i}| <1$, the eigenvalues are negative, and we conclude
that the points $(\text{F}_{1,2})$ are sinks.

\noindent\textbf{The case $(\text{F}_{\mathrm{A,B}})$}: 
As the fixed point $(\text{F}_{\mathrm{A}})$ lies on $\UU_{\mathrm{A}}$, 
one eigenvector of the linearization of the system at $(\text{F}_{\mathrm{A}})$
is tangential to $\UU_{\mathrm{A}}$.
The associated eigenvalue is $(n-1)/n$, which follows when
we set $A=0$ in~\eqref{eq:P} and compute
$P^{-1} P^\prime|_{(\text{F}_{\mathrm{A}})} = 1 -(1/n)$. 
There exists a second eigenvector which is
transversal to $\UU_{\mathrm{A}}$;
the associated eigenvalue is
$(-1/n)$, since $A^{-1} A^\prime |_{(\text{F}_{\mathrm{A}})} = -(1/n)$.
We conclude that $(\text{F}_{\mathrm{A}})$ is a saddle, 
so that generic orbits in $\SS$ do not tend to $(\text{F}_{\mathrm{A}})$.
However, there exists exactly one orbit that converges to $(\text{F}_{\mathrm{A}})$
along the stable subspace as $\tau\rightarrow \infty$.
Finally, since $\UU_{\mathrm{A}}$ coincides with the unstable manifold,
$(\text{F}_{\mathrm{A}})$ is a repellor in $\UU_{\mathrm{A}}$.
A similar analysis applies to $(\text{F}_{\mathrm{B}})$. 

\noindent\textbf{The case $(\text{F}_*)$}:
Let $J_*$ denote the linearization matrix of the system~\eqref{eq:evol} at $(\text{F}_*)$,
i.e., the Jacobian of the right hand side of~\eqref{eq:evol}.
The tangent space $\text{T}_{(\text{F}_*)}\SS$ of the constraint manifold $\SS$ at $(\text{F}_*)$ in $\XX$ is spanned by 
the vectors $(n,-m,0,0)^t,(0,0,n,-m)^t$. Let
$$
S = \begin{pmatrix} n & 0 \\
                   -m & 0 \\
                    0 & n \\
                    0 & -m
    \end{pmatrix} 
$$
and denote by $S^{-1}$ any $(2\times 4)$-matrix that is left-inverse w.r.t.\ $S$.
The restriction of $J_*$ to $\text{T}_{(\text{F}_*)}\SS$ is given by
$S^{-1} J_* S$, which yields
\begin{equation}\label{linat*}
J_* \big|_{\text{T}_{(\text{F}_*)}\SS} = S^{-1} J_* S = \frac{1}{m+n}\,
\begin{pmatrix}
0 & 1 \\
-2 (m+n-1) & m+n-1 
\end{pmatrix}\:.
\end{equation}
The eigenvalues of this matrix are
\begin{equation}
\lambda^*_{1,2} = \frac{m+n-1}{2 (m+n)}\, \pm \frac{\sqrt{(m+n-1)(m+n-9)}}{2 (m+n)}\:.
\end{equation}
We distinguish three qualitatively different cases:
when $(m+n) < 9$, there exists a non-vanishing imaginary part;
when $(m+n) = 9$, $\lambda_1^* = \lambda_2^*$;
when $(m+n) > 9$, the eigenvalues are real and $1> \lambda_1^* > \lambda_2^* > 0$.
For all $(m,n)$, the real part of $\lambda^*_{1,2}$ is positive,
which entails that $(\text{F}_*)$ is a repellor, i.e., locally stable
toward the past.

\subsection{Global dynamics}
\label{globaldynamics}

The global properties of the flow on the boundaries of $\SS$ are simple: 
since $\UU_{\mathrm{A}}$ and $\UU_{\mathrm{B}}$ are one-dimensional, the local stability analysis
implies the global dynamics.
$\UU_{\mathrm{A}}$ can be viewed as an interval, whose end points
are the stable fixed points $(\text{F}_1)$ and $(\text{F}_2)$.
In between there exist one additional fixed point, the repellor $(\text{F}_{\mathrm{A}})$.
It follows that every orbit (different from the fixed points) 
originates from $(\text{F}_{\mathrm{A}})$ and ends in $(\text{F}_{1})$ or $(\text{F}_2)$.
The picture on $\UU_{\mathrm{B}}$ is analogous.

On $\SS$, we observe that $P$ is monotone when $P\leq 0$, i.e., $P^\prime < 0$ when $P\leq 0$, 
which is because $m P^2 + n Q^2 \leq  1$.
Analogously, $Q^\prime <0$ when $Q \leq 0$ (which in turn corresponds to
$P^\prime > 0$ when $P\geq (1/m)$ via~\eqref{eq:trace}).
The flow on $\SS$ is thus particularly simple for $P\leq 0$ and $P\geq (1/m)$.

Now consider the non-negative function $Z = A^m B^n$ on $\bar{\XX}$
($Z$ is essentially the inverse of the rescaled volume density of the spatial metric).
Under the side-condition $C_2 = 0$, cf.~\eqref{eq:Ham}, hence in particular on $\bar{\SS}$, 
$Z$ attains a global maximum, namely at the point $(\text{F}_*)$.
On the boundaries $\UU_{\mathrm{A}}$ and $\UU_{\mathrm{B}}$, $Z$ becomes zero, which is its minimal value.
From~\eqref{eq:evol} we obtain that
\begin{equation}\label{eq:Zprime}
Z^\prime = Z \, [ 1 -(m+n) ( m P^2 + n Q^2) ]
\end{equation}
on $\bar{\SS}$.
Since $( m P^2 + n Q^2) > 1/(m+n)$ on $\bar{\SS}$
unless $P =Q =1/(m+n)$, 
the bracket in~\eqref{eq:Zprime} is negative almost everywhere.
In the special case $P =Q =1/(m+n)$ we obtain $Z^{\prime\prime} = 0$
and 
\begin{equation}
Z^{\prime\prime\prime} = -2 Z (m+n) (m+n-1)^2 \left[ m \left(A^2 - (\textfrac{1}{m+n})^2\right)^2
+ n \left(B^2 - (\textfrac{1}{m+n})^2\right)^2 \right]\:,
\end{equation}
which is non-positive and vanishes only when $A = B = 1/(m+n)$, i.e., at the fixed point $(\text{F}_*)$.
We infer that 
$Z$ is strictly monotonically decreasing along all orbits in $\SS \backslash (\text{F}_*)$.
This excludes that there exist any nontrivial periodic orbits in $\SS$ and
allows us to invoke the monotonicity principle:
the $\alpha$-limit of every orbit in $\SS$ must coincide with the fixed point
$(\text{F}_*)$;
the $\omega$-limit of every orbit must lie on the union of $\UU_{\mathrm{A}}$ and $\UU_{\mathrm{B}}$,
i.e., $A B \rightarrow 0 $ as $\tau\rightarrow \infty$.

Summarizing, we have proved the following theorem:
\begin{thm}[Global dynamics]\label{mm1thm} 
Let $k_g = k_h = -1$ and $m,n >1$ and consider the dynamical system~\eqref{eq:evol} on the two-dimensional
physical state space $\bar{\SS}$.
\begin{enumerate} 
\item 
Consider an orbit with $A = 0$, $B \neq 0$ 
that is different from $(\mathrm{F}_{\mathrm{A}})$.
The $\alpha$-limit of the orbit is the fixed point $(\mathrm{F}_{\mathrm{A}})$,
the $\omega$-limit is $(\mathrm{F}_{1})$ or $(\mathrm{F}_2)$.
The statement for $A\neq 0$, $B=0$ is analogous.
\item 
Consider the family of orbits with $A B \neq 0$ that are different from $(\mathrm{F}_*)$.
The $\alpha$-limit of every orbit is the fixed point $(\mathrm{F}_*)$.
There is one orbit whose $\omega$-limit is $(\mathrm{F}_{\mathrm{A}})$, another
one whose $\omega$-limit is $(\mathrm{F}_{\mathrm{B}})$.
The $\omega$-limit of a generic orbit is 
one of the equilibrium points $(\mathrm{F}_{1})$, $(\mathrm{F}_2)$. 
\end{enumerate} 
\end{thm}

\begin{figure}[Ht]
\begin{center}
\psfrag{F1}[cc][cc][1][0]{$(\text{F}_{1})$}
\psfrag{F2}[cc][cc][1][0]{$(\text{F}_{2})$}
\psfrag{FA}[lc][lc][1][0]{$(\text{F}_{\mathrm{A}})$}
\psfrag{FB}[rc][rc][1][0]{$(\text{F}_{\mathrm{B}})$}
\psfrag{B}[cc][cc][1][0]{$B$}
\psfrag{A0}[cc][cc][0.8][-62]{$A=0$}
\psfrag{B0}[cc][cc][0.8][-62]{$B=0$}
\subfigure[$m+n< 9$ ($D<10$)]{\label{mm1small}\includegraphics[height=0.35\textheight]{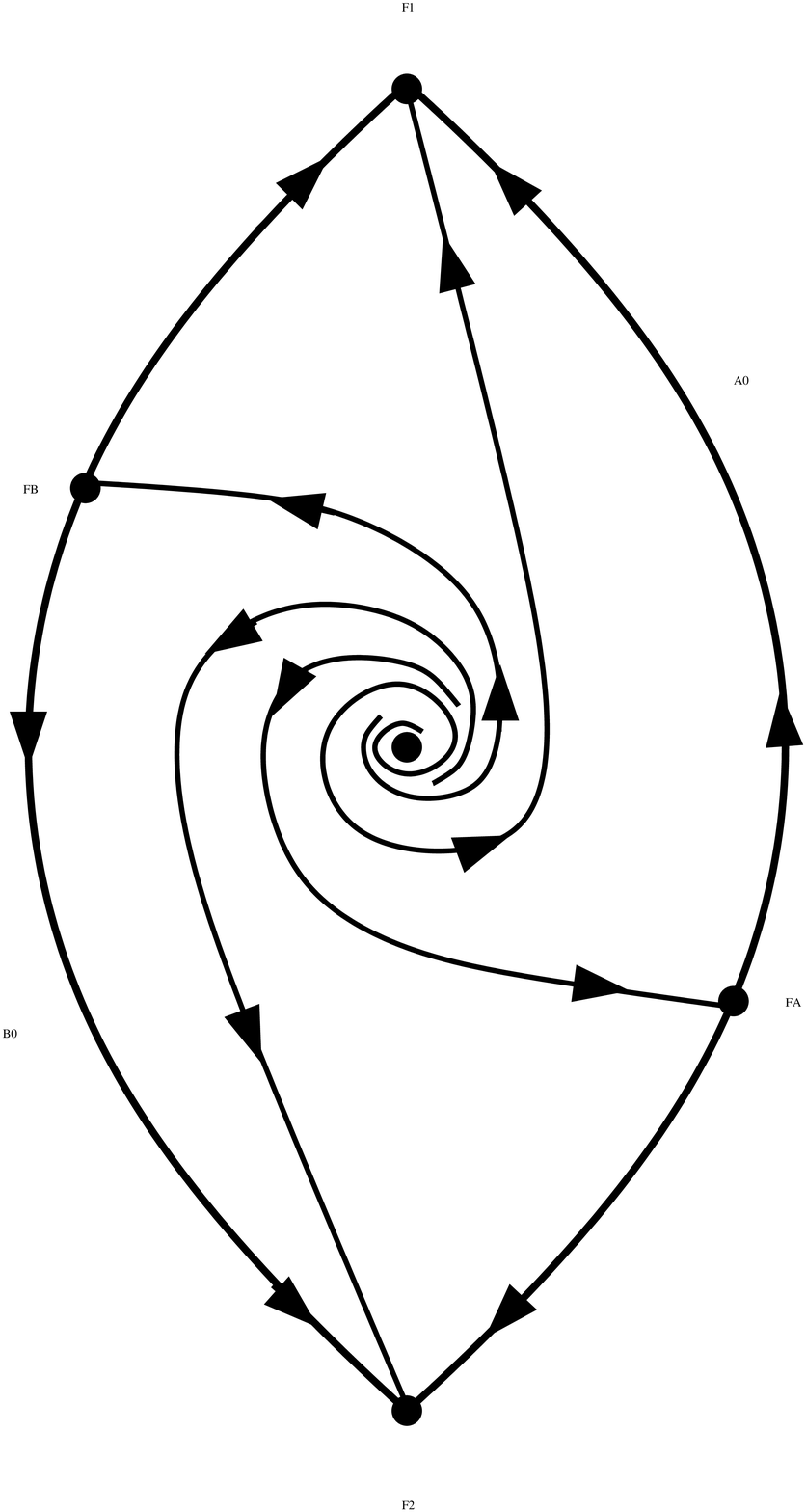}}
\qquad\qquad\qquad
\subfigure[$m+n > 9$ ($D>10$)]{\label{mm1big}\includegraphics[height=0.35\textheight]{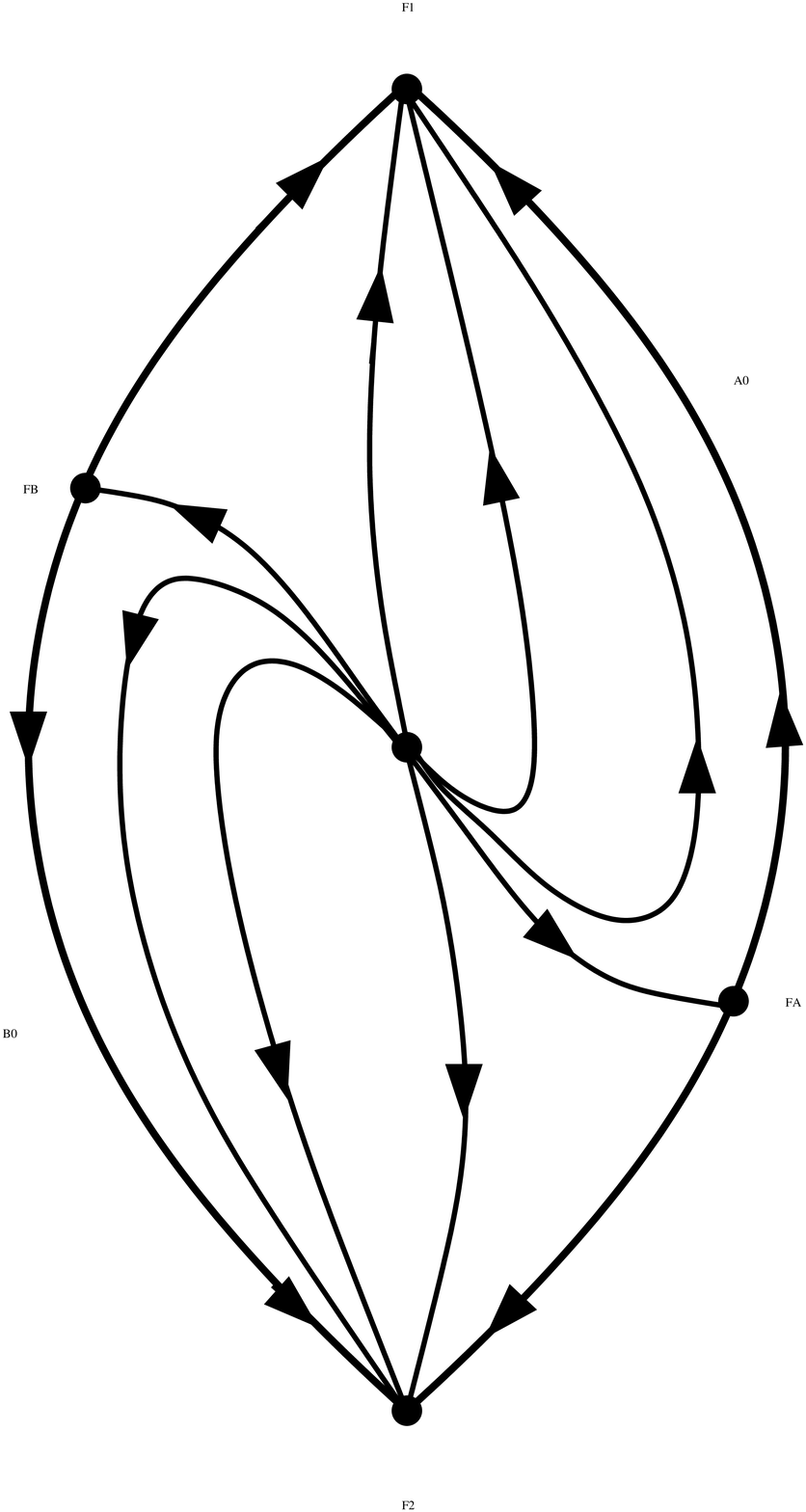}}
\caption{Schematic of the flow on the state space $\bar{\SS}$ for the cases $m+n < 9$
and $m+n > 9$. The point at the center is the fixed point $(\text{F}_{*})$. (The flow in the case $m+n=9$ 
looks qualitatively like (b); however, the eigenvectors of the linearization at $(\text{F}_{*})$ coincide.)}
\label{mm1fig}
\end{center}
\end{figure}

The flow of the dynamical system on $\bar{\SS}$ is depicted in Fig.~\ref{mm1fig}.

The behavior of the decoupled variable $H$ is given by Eq.~\eqref{eq:H-evol}: 
the equation implies that $H$ is monotonic in $\tau$.
By virtue of the inequalities $1 \geq ( m P^2 + n Q^2) \geq 1/(m+n)$ on $\bar{\SS}$
we obtain $H \leq  H^\prime \leq H/(m+n)$;
in particular,
$H \searrow -\infty$ as $\tau \nearrow +\infty$,
and $H \nearrow 0$
as $\tau \searrow -\infty$.

\begin{remark}
In~\cite{Dancer/Wang:1999,Dancer/Wang:2001} the $D$-dimensional Einstein vacuum equations 
for (Riemannian) metrics $d t^2 +a^2 g + b^2 h$ 
are analyzed
as a constrained Hamiltonian system, in particular as regards the integrability of the equations. 
The authors provide evidence that the cases $D=10,11$ are special (and perhaps integrable); 
indeed, at least for three subcases conserved quantities for the flow are constructed in~\cite{Dancer/Wang:1999}, 
see also the remark in Sec.~\ref{acceleration}. 
\end{remark}

\subsection{Asymptotics of the scale factors \boldmath $a$ and $b$ \unboldmath} \label{sec:solutions}

\quad\newline
Every orbit in $\SS$ gives rise to a solution $-dt^2 +a^2 g + b^2 h$ 
of the $D$-dimensional Einstein vacuum equations 
via the transformation~\eqref{vartrans}.
The particular asymptotic behavior of the orbit translates to characteristic
asymptotic behavior of the scale factors $a$ and $b$:

\begin{prop}[Asymptotic behavior]\label{asybehave}
An orbit that converges
to $(\mathrm{F}_{1,2})$ as $\tau\rightarrow \infty$ generates a solution with
asymptotic behavior of Kasner type, i.e.,
$a \sim t^p$, $b\sim t^q$ as $t\searrow 0$, where $m p + n q =1$ and $m p^2 + n q ^2 =1$.
When an orbit converges to $(\mathrm{F}_{*})$ as $\tau \rightarrow -\infty$, then
the corresponding solution is of Friedmann type as $t\rightarrow \infty$: $a \sim t$, $b\sim t$.
Finally, convergence to $(\mathrm{F}_{\mathrm{A,B}})$ as $\tau\rightarrow \infty$ leads to
$a \rightarrow \mathrm{const}$, $b\sim t$ and $a \sim t$, $b\rightarrow \mathrm{const}$ as $t\searrow 0$, respectively.
\end{prop}

\begin{remark}
Note that the orbits on the boundaries converge to $(\text{F}_{\mathrm{A,B}})$ as $\tau\rightarrow -\infty$.
The corresponding solutions satisfy $a \rightarrow \mathrm{const}$, $b\sim t$ and $a \sim t$, $b\rightarrow \mathrm{const}$ 
as $t\rightarrow \infty$.
\end{remark}

In the following we establish the above proposition; furthermore we analyze in more detail
the approach to Friedmann as $t\rightarrow \infty$.

\noindent\textbf{The case $(\text{F}_{1,2})$}:
Consider a solution that converges to $(\text{F}_1)$
as $\tau\rightarrow \infty$.
We have $H\rightarrow -\infty$, which suggests that the limit corresponds
to a singularity. Since $d t/d\tau = H^{-1}$ and $H$ decreases exponentially with $\tau$,
we find that $\tau\rightarrow \infty$ corresponds to $t\searrow 0$.

In a neighborhood of the fixed point 
$(\text{F}_1)$ 
the dynamical system~\eqref{eq:evol} on $\bar{\SS}$ can be approximated by its linearization at $(\text{F}_1)$,
which is given by
\begin{equation}\label{linatF1}
A^\prime = A [P_1 -1]\,, \qquad 
B^\prime = B [Q_1 -1]\,, \qquad
P \equiv P_1\,, \qquad 
Q\equiv Q_1 \:,
\end{equation}
cf.~the previous stability analysis.
It is important to note that 
the equations~\eqref{linatF1} coincide with the system of equations~\eqref{eq:evol}~\&~\eqref{eq:con}
obtained for $k_g=k_h =0$.
We conclude that all solutions that converge to $(\text{F}_1)$ as $\tau\rightarrow \infty$
behave asymptotically like
solutions representing a spacetime where both factors $g$ and $h$ are Ricci flat.
In other words, the dynamical
effects of curvature become negligible at the singularity.

When we solve~\eqref{linatF1} and the decoupled equation~\eqref{eq:H-evol} for $H$,
and recall that $d t/d\tau = H^{-1}$, we obtain via~\eqref{vartrans}:
\begin{equation}\label{Kasnertype}
a = a_0 \: t^{P_1} \,,\qquad b = b_0\: t^{Q_1} 
\end{equation}
for $t\searrow 0$; $a_0$ and $b_0$ are constants.
By construction, $m P_1 + n Q_1 =1$ and $m P_1^2 + n Q_1^2 =1$; $P_1 > 0$.
A completely analogous result is obtained for solutions that approach
$(\text{F}_2)$, where $(P_1,Q_1)$ is replaced by $(P_2,Q_2)$.

\noindent\textbf{The case $(\text{F}_*)$}:
Consider a solution that converges to $(\text{F}_*)$ as $\tau\rightarrow -\infty$.
We have $H\rightarrow 0$ and $t\rightarrow \infty$, 
which suggests that $(\text{F}_*)$ corresponds 
to the limit of infinite expansion.
Since $P=Q$ in the limit we expect Friedmann like behavior.

Assume that $m+n < 9$. 
In a neighborhood of the fixed point 
$(\text{F}_*)$ 
the dynamical system~\eqref{eq:evol} on $\SS$
is approximated by its linearization~\eqref{linat*}.
Using the variable transformation 
\begin{equation}
B= \frac{1}{\sqrt{9-(m+n)}} 
\begin{pmatrix}
1 & \sqrt{m+n-1} \\
0 & 4 \sqrt{m+n-1} 
\end{pmatrix}
\end{equation}
in $\text{T}_{(\text{F}_*)}\SS$, the linearized system takes a normal form
given by
\begin{equation}\label{Jstarnormal}
B^{-1} \Big( J_* \big|_{\text{T}_{(\text{F}_*)}\SS} \Big)\, B  =  
\frac{\sqrt{m+n-1}}{2 (m+n)} 
\begin{pmatrix}
\sqrt{m+n-1} & \sqrt{9-(m+n)} \\
-\sqrt{9-(m+n)} & \sqrt{m+n-1}
\end{pmatrix}\:.
\end{equation}

Based on this we can show that, modulo terms of order $O(t^{-(m+n-2)})$,
\begin{subequations}\label{Fstarapprox}
\begin{align}
\nonumber 
a & = t \Big[ 1 - n c_0 \sqrt{9-(m+n)}\, t^{-(m+n-1)/2} \cos(\cdot) \: + \\ 
& \qquad \qquad\qquad + n c_0\sqrt{m+n-1}\, t^{-(m+n-1)/2}  \sin(\cdot) \Big] \\
\nonumber 
b & = t \Big[ 1 - m c_0 \sqrt{m+n-1}\,   t^{-(m+n-1)/2} \sin(\cdot) \: + \\
& \quad\qquad\qquad + m c_0 \sqrt{9-(m+n)}\, t^{-(m+n-1)/2} \cos(\cdot) \Big] \:,
\end{align}
\end{subequations}
where the argument of the trigonometric functions is
\begin{equation}
(\cdot) = \left(-\textfrac{1}{2} \sqrt{m+n-1} \sqrt{9-(m+n)} \log (t/d_0)\right) \:;
\end{equation}
$c_0$, $d_0$ are constants.

When $m+n  \geq 10$ the eigenvalues $\lambda^*_{1,2}$ of the linearization of the dynamical system
at $(\text{F}_*)$ are real with $1> \lambda_1^* \geq  2 \lambda_2^* > 0$.
Oscillatory terms as in~\eqref{Fstarapprox} do not occur in this case.
A thorough analysis of the system yields that orbits converging to $(\text{F}_*)$ as
$\tau\rightarrow -\infty$ correspond to solutions of the type
\begin{subequations}\label{Fstarapprox2}
\begin{align}
a & = t \left[ 1 + n c_0 t^{-\lambda^*} + O(t^{-2 \lambda^*}) \right] \\
b & = t \left[ 1 - m c_0 t^{-\lambda^*} + O(t^{-2 \lambda^*}) \right] 
\end{align}
\end{subequations}
as $t\rightarrow \infty$, where $\lambda^* = (m+n) \lambda^*_2 = \frac{1}{2} ( m+n-1 -\sqrt{(m+n-1)(m+n-9)} )$.
(In~\eqref{Fstarapprox2} there appears only one free parameter; 
the second parameter of the two-parameter family of solutions 
is connected to terms of the order $t^{-(m+n) \lambda^*_1}$.)

In the special case $m+n =9$ 
the linearization of the dynamical system
at $(\text{F}_*)$ is represented by a non-diagonalizable matrix.
Orbits converging to $(\text{F}_*)$ as
$\tau\rightarrow -\infty$ correspond to solutions of the type
\begin{subequations}\label{Fstarapprox3}
\begin{align}
a & = t \left[ 1 + n c_0 (\log t) t^{-4} + n d_0 t^{-4} + O\big( (\log t)^2 t^{-8} \big)  \right] \\
b & = t \left[ 1 - m c_0 (\log t) t^{-4} -m d_0 t^{-4} +  O\big( (\log t)^2 t^{-8} \big)\right]
\end{align}
\end{subequations}
as $t\rightarrow \infty$, where $c_0$, $d_0$ are constants.

\noindent\textbf{The case $(\text{F}_{\mathrm{A,B}})$}: 
Consider a solution along the orbit 
that converges to $(\text{F}_{\mathrm{A}})$ as $\tau\rightarrow \infty$.
We have $H\rightarrow -\infty$ and $t\rightarrow 0$ as $\tau\rightarrow \infty$, 
which suggests that the limit corresponds to a singularity.
Using again approximation techniques we eventually find that 
\begin{equation}
a \rightarrow a_0 \,\qquad b = \frac{\sqrt{m+n-1}}{\sqrt{n-1}}\, t 
\end{equation}
as $t\rightarrow 0$. This completes the proof of Proposition~\ref{asybehave}.

\begin{coro}
Combining the above results with the statements on the behavior of generic orbits in $\SS$,
cf.~Theorem~\ref{mm1thm}, we find that generic solutions of the
$D$-dimensional vacuum equations
are of Kasner type, as in~\eqref{Kasnertype}, as $t\rightarrow 0$
and of Friedmann type, as in~\eqref{Fstarapprox},~\eqref{Fstarapprox2}, or~\eqref{Fstarapprox3}, as $t\rightarrow \infty$.
Interestingly enough, the approach to the Friedmann solution is oscillatory if and only if
$m+n < 9$ ($D<10$).
\end{coro}

\section{AVTD condition} 
\label{sec:AVTD} 

In this section we briefly address the question of whether there exist
families of perturbations (that are general in the sense that they depend on the maximal number
of free functions) of the $D$-dimensional models
described above that exhibit quiescent behavior at the singularity, 
cf.~\cite{demaret:etal:1985, damour:etal:kasnerlike}.

Consider a $(d+1)$-dimensional Kasner spacetime, with Kasner exponents 
$p_i$, $i=1\dots d$, and line element 
\begin{equation}
-dt^2 + \sum_{i=1}^d t^{2p_i} (dx^i)^2\:.
\end{equation}
The vacuum Einstein equations imply the Kasner relations 
\begin{equation}\label{kaskas}
\sum_{i=1}^d p_i = 1, \qquad \sum_{i=1}^d p_i^2 = 1\:.
\end{equation}
We consider the case where $d = m+n$, with $p_i = P$, $i=1,\dots m$, and $p_i
= Q$, $i=m+1\dots m+n$. Then the Kasner relations~\eqref{kaskas} read 
\begin{equation}\label{eq:genkas} 
mP + nQ = 1, \qquad  mP^2 + nQ^2 = 1\:,
\end{equation} 
which we recognize as the constraint equations~(\ref{eq:con}) in case $A = B = 0$. 
These equations hold at the equilibrium points $(\text{F}_{1,2})$, cf.~\eqref{F12}.

Following \cite{demaret:etal:1985}, 
see also \cite[\S 3]{damour:etal:kasnerlike}, 
the condition
for asymptotically velocity-term dominated (AVTD) behavior for an 
asymptotically Kasner spacetime is 
$$
1 + p_1 -p_d - p_{d-1} > 0\:,
$$
which in terms of the generalized Kasner exponents $P,Q$ reads 
\begin{equation}\label{eq:vcond} 
1 + P - 2Q > 0\:.
\end{equation} 
At the equilibrium points $(\text{F}_{1,2})$, we have 
\begin{equation}
P_{1,2} = \frac{1}{m+n}\left ( 1 \pm n \sqrt{\frac{m+n-1}{mn}} \right ) ,\quad
Q_{1,2} = \frac{1}{m+n} \left ( 1 \mp m \sqrt{\frac{m+n-1}{mn}} \right ).
\end{equation}
The condition~\eqref{eq:vcond} at $(\text{F}_{1})$, i.e., $1 + P_1 -2 Q_1 > 0$, 
is satisfied for all $(m,n)$ such that 
\begin{equation}\label{AVTD1cond}
m > \frac{4n}{n-1} \:, \qquad n >1 \:.
\end{equation}
Equivalently, $1 + P_2 -2 Q_2 > 0$, when $m>1$ and $n > 4 m/(m-1)$.
Eq.~\eqref{AVTD1cond} implies, with $d=m+n$,
\begin{equation}
d > \frac{n(n+3)}{n-1} \qquad (n>1)\:.
\end{equation}
For $n=2,3,4,5,\ldots$ the right hand side is $10,9,9\frac{1}{3}, 10, \ldots$, and 
for $n \geq 3$, the right hand side is monotonically increasing. 
Therefore we find, in agreement with the result of
\cite{demaret:etal:1985}, that $d \geq 10$ is a necessary condition for AVTD behavior. \\

\section{Accelerating cosmologies from compactification} 
\label{acceleration}

Dimensional (Kaluza-Klein) reduction transforms 
classes of $D$-dimensional vacuum spacetimes 
into classes of $(1+m)$-dimensional spacetimes (where typically $m=3$)
with nonlinear scalar fields.
In some instances the resulting $(1+m)$-dimensional models
represent accelerating cosmologies, as has been demonstrated
by a sizable number of examples, see, e.g., \cite{Townsend/Wohlfarth:2003, Wohlfarth:2004, Ohta:2005} 
and~\cite{Chen/Ho/Neupane/Ohta/Wang:2003, Chen/Ho/Neupane/Wang:2003}. 
In the following we present a systematic treatment of accelerating cosmologies arising from the
compactification of the $D$-dimensional spacetimes discussed in Sec.~\ref{sec:analysis}.
In particular, we prove existence and uniqueness of a model that exhibits eternal acceleration. 

Consider the family of $D = 1+m +n$ dimensional vacuum spacetimes $\mathbb{R}\times M \times N$ 
with metrics $-d t^2 + a^2 g + b^2 h$ that
has been constructed.
We perform a dimensional reduction, i.e., we assume that the
$m$ spatial dimensions connected to the factor $g$ are 
the large spatial dimensions representing the classical spacetime (where typically $m=3$
in order to obtain a four-dimensional spacetime), while
the $n$ spatial dimensions connected to $h$ are to be compactified;
\begin{equation}
-d t^2 + a^2 g + b^2 h\:  \stackrel{\mbox{\scriptsize\text{reduction}}}{\longrightarrow}\:\,\gamma = e^{-2 \phi}  (-d t^2 + a^2 g)\:.
\end{equation}
We choose a conformal Einstein frame for
the $1+m$ metric, see, e.g.,~\cite[App.~A]{Chen/Ho/Neupane/Wang:2003}, i.e., we conformally rescale $-dt^2 +a^2 g$ by choosing $\phi$
according to 
\begin{equation}
e^\phi = b^{- n/(m-1)}\:,
\end{equation}
which entails that the reduction of the $(1+m+n)$-dimensional Einstein-Hilbert action results in an effective
action, again of Einstein-Hilbert type, for the metric $\gamma$ and a nonlinear scalar field $\varphi$
that stems from the compactified dimensions, 
$\varphi = -(8 \pi)^{-1/2}\, n^{-1/2} (m-1)^{1/2} (m+n-1)^{1/2}\, \phi$. 
Therefore, $(M,\gamma,\varphi)$ satisfies the Einstein nonlinear scalar field equations, i.e.,
\begin{equation}
\Ric_\gamma - \textfrac{1}{2} \mathrm{Scal}_\gamma\, \gamma = 8 \pi\, T \:,
\end{equation}
where $T$ is the energy-momentum tensor of the scalar field,
\begin{equation}
T_{\mu\nu} = \nabla_\mu \varphi \nabla_\nu \varphi - 
\left[\frac{1}{2} \nabla^\sigma \varphi \nabla_\sigma \varphi + V(\varphi) \right]\: \gamma_{\mu\nu} \qquad
(\mu,\nu = 0 \ldots m)\:.
\end{equation}
The 
potential $V(\varphi)$ of the scalar field is an exponential function,
\begin{equation}\label{exppot}
V(\varphi) = (8 \pi)^{-1} \,\frac{n (m + n -1)}{2} 
\exp \left[ - \sqrt{8 \pi} \frac{2}{\sqrt{m-1}} \sqrt{\frac{m+n-1}{n}} \: \varphi\right]\:.
\end{equation}
For details on dimensional (Kaluza-Klein) reduction see, e.g.,~\cite{Heinzle/Rendall:2005}.

For the metric $\gamma$, by
introducing a new time variable $\bar{t}$ through
$d \bar{t}/ d t = b^{n/(m-1)}$
we finally obtain
\begin{equation}
\gamma = b^{2 n/(m-1)} (-d t^2 + a^2 g) = -d \bar{t}^2 + \left( a b^{n/(m-1)} \right)^2\, g = -d\bar{t}^2 + \bar{a}^2\, g 
\end{equation}
with $\bar{a} = a b^{n/(m-1)}$. 

In the following we investigate 
the spacetimes $\mathbb{R}\times M$ with metrics $\gamma$; in particular we 
focus on the question of whether these solutions give rise to
cosmological models that exhibit accelerated expansion.

By construction, every orbit in $\SS$ is associated with a solution $\gamma = -d \bar{t}^2 + \bar{a}^2 g$ 
of the $(1+m)$-dimensional Einstein equations (with nonlinear scalar field $\phi$), where
the scale factor $\bar{a} = a b^{n/(m-1)}$ is determined via~\eqref{vartrans}.
A simple calculation shows that the derivative of the scale factor $\bar{a}$ is 
\begin{equation}\label{baraone}
\frac{d \bar{a}}{d \bar{t}} = \frac{1}{m-1} \, \frac{1}{A}\: [1- P]\:.
\end{equation}
Since $P<1$ on the entire state space $\bar{\SS}$, $d\bar{a}/d \bar{t}$ is positive
for all solutions, so that
all metrics $\gamma$ describe expanding cosmologies.

Differentiating~\eqref{baraone} we obtain
\begin{equation}\label{baratwo}
\frac{d^2 \bar{a}}{d\bar{t}^2} = \frac{a b^{-n/(m-1)}}{m-1} H^2  
\Big[-(1-P)^2 - (m+n-1) \left(k_g (m-1) A^2 + k_h n B^2\right)\Big]\:.
\end{equation}
The first term in brackets is negative, the second term is non-negative on the state space $\bar{\SS}$.
Clearly, in a neighborhood of the fixed points $(\text{F}_{1,2})$, where $A=B=0$,
the sum is negative, hence $d^2 \bar{a}/d\bar{t}^2 < 0$. However, in the following
we will establish the existence of a domain $\AA \subseteq \SS$ such that
\begin{equation}
\frac{d^2 \bar{a}}{d\bar{t}^2} > 0 \quad \text{for all } (A,B,P,Q) \in \AA\subseteq \SS\:.
\end{equation}
The domain $\AA$ is the \textit{domain of acceleration}:
whenever an orbit passes through this domain, the cosmological model $(\mathbb{R}\times M, \gamma)$ it represents
undergoes accelerated expansion.

\begin{remark}
As before, in~\eqref{baraone}, $k_g = k_h =-1$ is understood. However, recall 
that setting $k_g=0$/$k_h =0$ corresponds to setting $A=0$/$B=0$;
thus the cases $k_g =0$, $k_h =-1$ 
and $k_g =-1$, $k_h =0$ are included in our treatment by our
investigation of the boundaries of $\SS$.
\end{remark}

Consider first the boundary $\UU_{\mathrm{A}}$ given by $A=0$ in $\bar{\SS}$. Setting $A=0$ in~\eqref{baratwo}
we find by simple algebraic manipulations of the bracket in~\eqref{baratwo} 
and by using the Hamiltonian constraint that 
\begin{equation}
\frac{d^2 \bar{a}}{d\bar{t}^2} \propto \frac{m+n-1}{k_{mn}} - 
\frac{k_{mn}}{n} \left( P - \frac{m+n}{k_{mn}} \right)^2
\end{equation}
with $k_{mn} = m(m+n) + n$. 
We infer that $d^2 \bar{a}/d\bar{t}^2 > 0$ in a neighborhood
of $P = (m+n)/k_{m n}$, namely for $P \in (\pi_-,\pi_+)$, where
\begin{equation}
\pi_\pm = \frac{m+n \pm n \sqrt{m+n-1}}{k_{mn}}\:;
\end{equation}
for all $m$ and $n$, $\pi_\pm$ are positive.
Hence, on the boundary $A=0$ there exists a non-empty interval of acceleration;
note that the fixed point $(\text{F}_{\mathrm{A}})$ is not an element of this interval.

\begin{remark}
Since the orbit $(\text{F}_{\mathrm{A}})$--$(\text{F}_{1})$ on the boundary $\UU_{\mathrm{A}}$ 
passes through an interval of acceleration, the corresponding model $(\mathbb{R}\times M, \gamma)$
exhibits a (finite) phase of accelerated expansion; note that $M$ is Ricci flat for this model.
This solution was originally found in~\cite{Townsend/Wohlfarth:2003}; it provided the first
example for accelerated expansion from compactification.
\end{remark}

On the boundary $\UU_{\mathrm{B}}$, by setting $B=0$ in~\eqref{baratwo} and using the constraint, we obtain
\begin{equation}
\frac{d^2 \bar{a}}{d\bar{t}^2} \propto - \frac{m}{n} (m+n-1) \left( P -\frac{1}{m}\right)^2\:.
\end{equation}
Hence, $d^2 \bar{a}/d\bar{t}^2 = 0$ at the fixed point $(\text{F}_{\mathrm{B}})$
and negative elsewhere on $B=0$.

\begin{figure}[Ht]
\begin{center}
\psfrag{F1}[cc][cc][1][0]{$(\text{F}_{1})$}
\psfrag{F2}[cc][cc][1][0]{$(\text{F}_{2})$}
\psfrag{FA}[lc][lc][1][0]{$(\text{F}_{\mathrm{A}})$}
\psfrag{FB}[rc][rc][1][0]{$(\text{F}_{\mathrm{B}})$}
\psfrag{F*}[cc][cc][1][0]{$(\text{F}_{*})$}
\psfrag{B}[cc][cc][1][0]{$B$}
\psfrag{A0}[cc][cc][0.8][-62]{$A=0$}
\psfrag{B0}[cc][cc][0.8][-62]{$B=0$}
\psfrag{A}[cc][cc][1.5][0]{$\AA$}
\psfrag{al}[cc][cc][0.8][0]{$\alpha_\psi$}
\subfigure[$m,n$ arbitrary]{\label{mm1accel}\includegraphics[height=0.35\textheight]{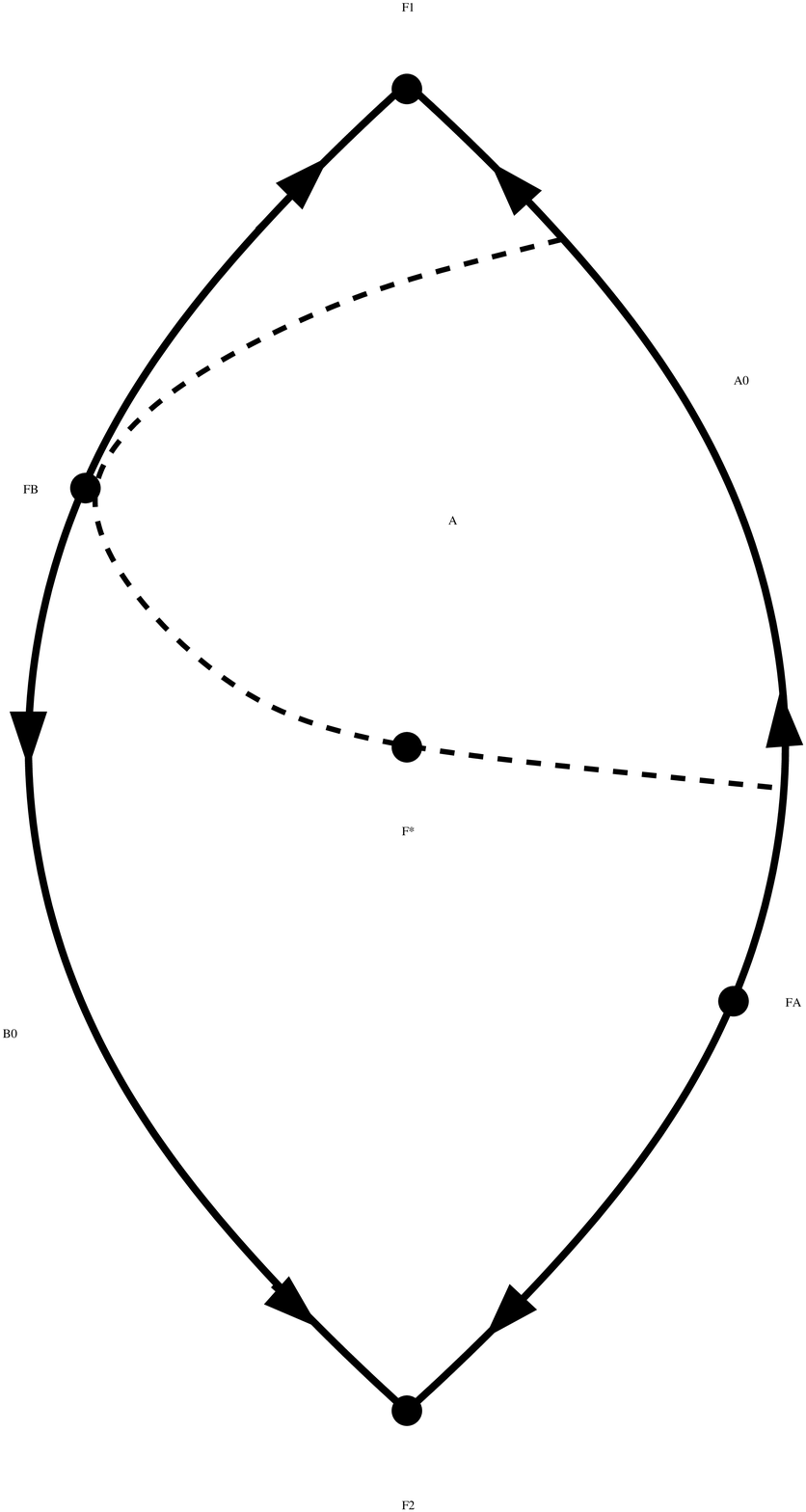}}\qquad\qquad\qquad
\subfigure[$m+n \geq 9$ ($D\geq 10$)]{\label{mm1unique}\includegraphics[height=0.35\textheight]{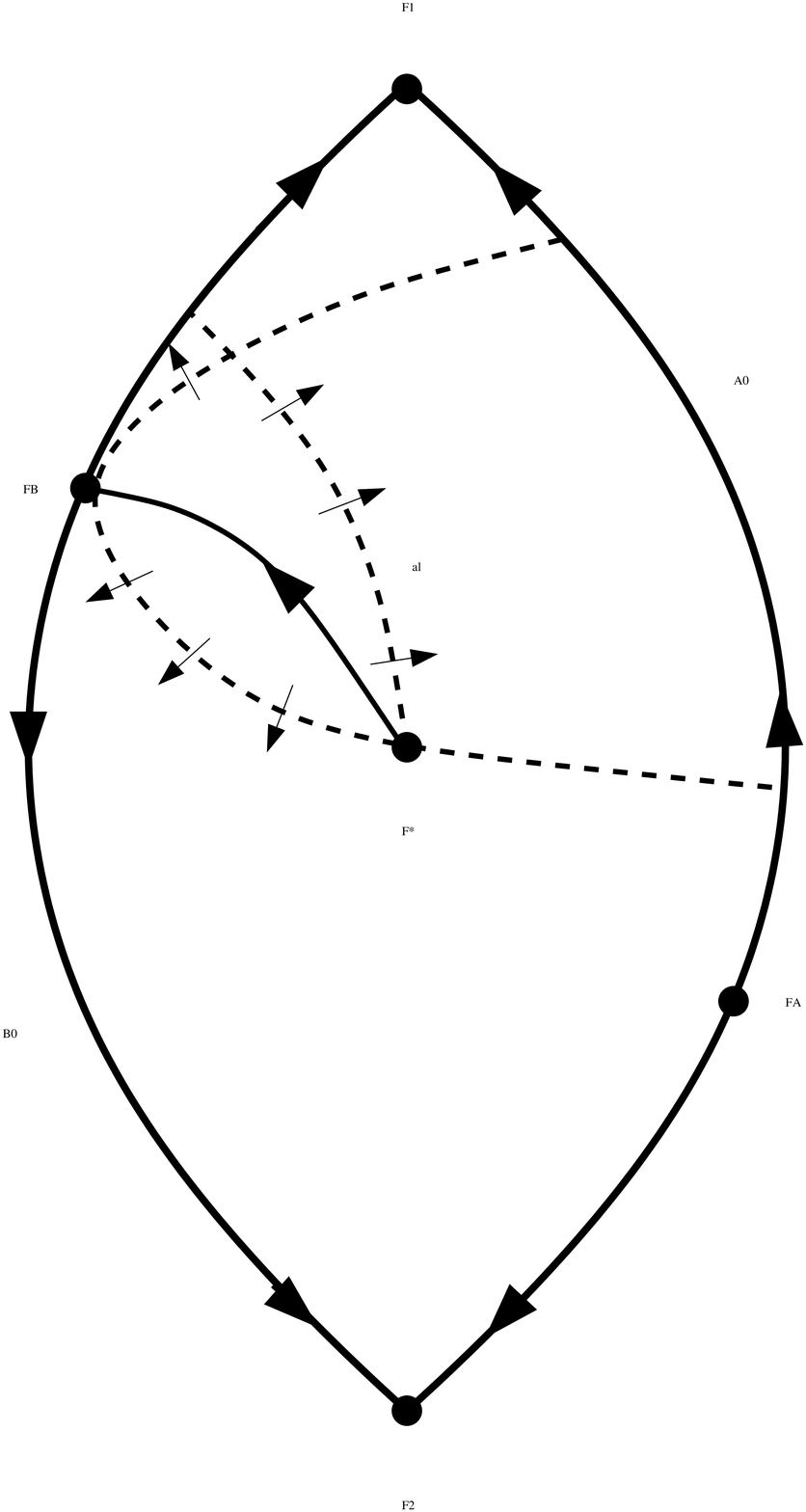}}
\caption{The domain of acceleration $\AA$ on the state space $\bar{\SS}$. Whenever an orbit passes 
through this domain, the cosmological model $(\mathbb{R}\times M, \gamma)$ it represents
undergoes accelerated expansion. For $m+n\geq 9$ ($m >2$) there exists a unique orbit that lies entirely in $\AA$;
the associated model exhibits eternal acceleration.}
\label{mm1accelall}
\end{center}
\end{figure}

Inserting the Hamiltonian constraint into Eq.~\eqref{baratwo} 
it is straightforward to show
that $\partial \AA$ (given by $d^2\bar{a}/d \bar{t}^2 = 0$) defines 
an ellipse in $\bar{\SS}$.
Combining this fact with the results collected above 
we conclude that 
the domain of acceleration $\AA$ in $\SS$ possesses the following main properties: 
$\bar{\AA}$ intersects the boundary $A = 0$ in an interval
that does not contain $(\text{F}_{\mathrm{A}})$ and 
the boundary $B=0$ in the point $(\text{F}_{\mathrm{B}})$; 
$\partial\AA$ is tangential to the boundary $B=0$ at that point.
Moreover, a simple calculation shows that 
$(\text{F}_{*})$ lies on $\partial\AA$; see Fig.~\ref{mm1accel}.

The boundary of $\AA$ consists of three semipermeable segments:
the flow of the dynamical system on the segment $(A=0)-(\text{F}_{*})$
is directed toward the interior of $\AA$, while the flow on
the segments $(\text{F}_{*})-(\text{F}_{\mathrm{B}})$ and
$(\text{F}_{\mathrm{B}})-(A=0)$ is directed toward the exterior. 
This is because
\begin{equation}
\frac{\partial}{\partial \tau} \left(\frac{d^2 \bar{a}}{d\bar{t}^2} \right) 
\Big|_{d^2 \bar{a}/d\bar{t}^2 =0} 
\propto  -\left( P -\frac{1}{m}\right)^2 \left(P -\frac{1}{m+n}\right)\:.
\end{equation}

Consider the case $m+n<9$ ($D < 10$). The cosmological model $(\mathbb{R}\times M, \gamma)$ 
possesses a phase of accelerated expansion whenever the associated orbit in $\SS$
passes through $\AA$.
Superimposing Fig.~\ref{mm1small} and 
Fig.~\ref{mm1accel} 
we see that a generic solution (represented by an orbit converging to
$(\text{F}_{1})$ or $(\text{F}_{2})$) 
does not exhibit accelerated expansion for early times.
However, toward the future, 
there exist infinitely many phases of accelerated expansion
(associated with the orbit spiraling out from $(\text{F}_{*})$).
The properties of the non-generic solution converging to $(\text{F}_{\mathrm{A}})$ are
analogous, but the solution converging to $(\text{F}_{\mathrm{B}})$
is of a different kind: it generates the single model that 
is accelerating for early times (i.e., in a neighborhood of the singularity).

Now consider the case $m+n\geq 9$ ($D\geq 10$), i.e., Fig.~\ref{mm1big} together with
Fig.~\ref{mm1accel}.
The figures suggest that, in contrast to the case $m+n <9$, there might exist
one orbit that lies entirely in $\AA$. We thus formulate the following

\begin{thm}[Existence and uniqueness of eternal acceleration]\label{eternalthm}
Let $m+n \geq 9$ (with $(m,n) \neq (2,7)$); $k_g = k_h =-1$. 
Then there exists a unique solution $\gamma = -d\bar{t}^2 + \bar{a}^2 g$
of the $(1+m)$-dimensional Einstein equations (with nonlinear scalar field $\varphi$) 
arising from the dimensional reduction of a $(1+m+n)$-dimensional vacuum solution,
such that $d^2 \bar{a}/d \bar{t}^2 > 0$ for all $\bar{t}$.
\end{thm}

\proof 
Since $(\text{F}_{1})$, $(\text{F}_{2})$, and $(\text{F}_{\mathrm{A}})$ 
do not lie in $\bar{\AA}$, orbits converging to any of these fixed points 
cannot generate models with eternal acceleration.
This leaves the orbit connecting $(\text{F}_{*})$ with $(\text{F}_{\mathrm{B}})$
as the only candidate.
To establish eternal acceleration for the associated model
we must show that this orbit lies entirely in $\AA$.
To that end consider the intersection of the plane
\begin{equation}
\left\{ 
(A,B,P,Q) \in \XX \:|\: \cos\psi \left(A - \frac{1}{m+n}\right) - \sin\psi \left(P - \frac{1}{m+n}\right) = 0 \right\} \;,
\end{equation}
where $\psi$ is a constant, and the state space $\bar{\SS} \subseteq \XX$. This intersection generates a curve $\alpha_\psi$ in $\bar{\SS}$ 
that connects $(\text{F}_{*})$ with a point on $B=0$, cf.~Fig.~\ref{mm1unique}.
When $\psi$ is chosen according to
\begin{equation}
\psi = \mathrm{arccot} \left(\frac{1}{2}\sqrt{m+n-1} \:\big[ \sqrt{m+n-1} + \sqrt{m+n-9} \,\big] \right) \:,
\end{equation}
then $\alpha_\psi$ possesses the following favorable properties:
(i) $\alpha_\psi$ connects $(\text{F}_{*})$ with a point on $B=0$ with $P > 1/m$, i.e.,
with a point on the $(\text{F}_{\mathrm{B}})-(\text{F}_{1})$ segment
(and $\alpha_\psi$ does not intersect the segment $(\text{F}_{*})-(\text{F}_{\mathrm{B}})$ of $\partial \AA$);
(ii) $\alpha_\psi$ is semipermeable for the flow of the dynamical system:
orbits can pass through $\alpha_\psi$ in one direction only (namely from the lower left side to
the upper right); see Fig.~\ref{mm1unique}. 
We have thus constructed a region $\AA_\psi\subseteq \AA$, enclosed by $\alpha_\psi$ and segments of $\partial A$, that
is past invariant for the flow of the dynamical system. Since the orbit converging 
to $(\text{F}_{\mathrm{B}})$ lies in $\AA_\psi$ as $\tau\rightarrow \infty$ 
it follows that it must lie in $\AA_\psi$ and thus in $\AA$ for all times. By this the theorem is established.

\begin{remark}
For one special case of $m$ and $n$, namely $m=2$, $n=7$, the above argument fails
(in particular (ii) does not hold).
A thorough numerical investigation yields that this is indeed because the statement is wrong:
the orbit connecting $(\text{F}_{*})$ with $(\text{F}_{\mathrm{B}})$ does not lie entirely in $\AA$ but 
intersects $\partial \AA$ 
transversally on the segment $(A=0)-(\text{F}_{*})$;
hence the orbit lies in $\AA$ for early times and outside for late times.
The numerical evidence thus suggests that in the special case $(m,n)=(2,7)$ eternal acceleration is impossible.
\end{remark}

\begin{remark}
In three special cases the ``orbit of eternal acceleration'' can be obtained explicitly:
for $(m,n) \in \{(2,8), (3,6), (5,5)\}$ the orbit is given simply as
\begin{equation}
\left\{ \left(\frac{1}{m}-\frac{1}{m+n}\right)  \left(A - \frac{1}{m+n}\right) -  
\left( A_{(\mathrm{F}_{\mathrm{B}})} -\frac{1}{m+n} \right) \left(P - \frac{1}{m+n}\right) = 0 \right\} \:,
\end{equation}
where $A_{(\mathrm{F}_{\mathrm{B}})} = (m-1)^{1/2} m^{-1} (m+n-1)^{-1/2}$, cf.~\eqref{FB}.
This is analogous to the analysis of~\cite{Dancer/Wang:1999}, in which it is shown that
the $D$-dimensional Einstein vacuum equations (for Riemannian metrics) regarded as a Hamiltonian system 
are integrable for $(m,n) \in \{(2,8), (3,6), (5,5)\}$. This coincidence 
suggests that the analysis of~\cite{Dancer/Wang:1999,Dancer/Wang:2001} might also apply in the Lorentzian case.
\end{remark}

\begin{cor}\label{fullcor}
The generic solutions cannot exhibit eternal acceleration.
Orbits that converge to $(\mathrm{F}_{1})$ correspond to solutions
that exhibit a phase of accelerated expansion $(t_{\mathrm{i}}, t_{\mathrm{o}})$; there are two subcases: 
solutions corresponding to orbits closer to the boundary $A=0$ possess
a finite phase of accelerated expansion, i.e., $0< t_{\mathrm{i}} < t_{\mathrm{o}} < \infty$, while
solutions corresponding to orbits closer to $B=0$ 
accelerate forever toward the future, i.e., $0< t_{\mathrm{i}} < t_{\mathrm{o}} = \infty$.
In contrast, accelerated expansion does not occur for all orbits that converge to $(\mathrm{F}_{2})$:
orbits closer to $A=0$ generate solutions with decelerating expansion.
(Note that the same holds for the non-generic solution converging to $(\mathrm{F}_{\mathrm{A}})$.)
The expansion for orbits closer to $B=0$ is accelerating in $(t_{\mathrm{i}}, t_{\mathrm{o}})$ with
$0< t_{\mathrm{i}} < t_{\mathrm{o}} = \infty$.
\end{cor}

The proof of the corollary is based on the established properties of the
orbit of eternal acceleration $(\text{F}_{*})$--$(\text{F}_{\mathrm{B}})$; $(m, n) \not\in \{(2,7),(7,2)\}$ is
necessary.

We conclude by discussing the leading term of the scale factor $\bar{a}(\bar{t})$
of the model $(\mathbb{R}\times M, \gamma)$ as $\bar{t} \rightarrow 0$ and
$\bar{t}\rightarrow \infty$.

\begin{description}
\item[\textbf{$(\text{F}_{1,2})$}]
For the generic orbit that approaches the Kasner fixed point $(\text{F}_{1})$
or $(\text{F}_{2})$ as $t\rightarrow 0$
we obtain $\bar{t} \propto t^{m(1-P)/(m-1)}$, hence $t\rightarrow 0$ corresponds
to $\bar{t}\rightarrow 0$ and we obtain
$\bar{a}\propto \bar{t}^{1/m}$ as $\bar{t} \rightarrow 0$.
\item[\textbf{$(\text{F}_*)$}] 
In the limit we have $\bar{t} \propto t^{(m+n-1)/(m-1)}$
and the leading term in $\bar{a}$ is proportional to $\bar{t}$.
\item[\textbf{$(\text{F}_{\mathrm{A,B}})$}] The orbit converging to $(\text{F}_{\mathrm{A}})$
generates a solution that satisfies $\bar{t} \propto t^{(m+n-1)/(m-1)}$ and
$\bar{a} \propto \bar{t}^{n/(m+n-1)}$ in the limit $\bar{t}\rightarrow 0$.
The solution corresponding to the orbit converging to $(\text{F}_{\mathrm{B}})$ 
satisfies $\bar{t} \propto t$ and $\bar{a} \propto \bar{t}$ in the limit $\bar{t}\rightarrow 0$.
\end{description}

\begin{remark}
The fixed point $(\text{F}_{\mathrm{A}})$ represents a model $-d\bar{t}^2 + \bar{a}^2 g$ where
$g$ is Ricci flat; the scale factor satisfies a power law, $\bar{a} \propto \bar{t}^{n/(m+n-1)}$.
This model is well-known, it is the simplest of solutions of the Einstein nonlinear scalar field 
equations with exponential potential~\eqref{exppot}. Since the exponent in $V(\varphi)$ is 
overcritical, the power in the scale factor $\bar{a}$ is less than one. \\
\end{remark}

\section{The case $k_g \geq 0$, $k_h \leq 0$} 
\label{mpsec}

Consider again the metric $-d t^2 + a^2 g + b^2 h$ with $\Ric_g = k_g (m+n-1) g$
and $\Ric_h = k_h (m+n-1) h$. 
Assume $k_g =1$ and $k_h= -1$.
For the analysis of this case
the equation system~\eqref{eq:evol} is ill-adapted since the 
variables become ill-defined (which is due to the possibility of $H$ going through zero).
We replace the system~\eqref{eq:evol}
by equations that are adapted to the case $k_g =1$, $k_h =-1$.
With 
\begin{equation}
D = \left( H^2 + k_g \frac{m(m+n-1)}{a^2}\right)^{1/2} 
\end{equation}
define, in close analogy to the previous definitions,
\begin{equation}
P =\frac{p}{D} \,,\quad
Q=\frac{q}{D} \,,\quad
A=(a D)^{-1}\,,\quad
B=(b D)^{-1}\:.
\end{equation}
This transforms the Einstein equations to a system consisting of
the decoupled equation
\begin{equation}\label{mp1D}
D^\prime = D \left( (m P + n Q)(m P^2 +n Q^2) + k_g m (m+n-1) P A^2 \right)
\end{equation}
and the coupled equations
\begin{subequations}\label{mp1}
\begin{align}
\label{mp1A}
A^\prime & = A \left[ P - (m P + n Q)(m P^2 +n Q^2) - k_g \mathcal{P} A^2 \right] \\
\label{mp1B}
B^\prime & = B \left[ Q - (m P + n Q)(m P^2 +n Q^2) - k_g \mathcal{P} A^2 \right] \\
\label{mp1P}
P^\prime & = P \left[ (m P + n Q)(1 - m P^2 - n Q^2) - k_g \mathcal{P} A^2 \right] + k_g (m+n-1) A^2 \\
\label{mp1Q}
Q^\prime & = Q \left[ (m P + n Q)(1 - m P^2 - n Q^2) - k_g \mathcal{P} A^2  \right] + k_h (m+n-1) B^2\:,
\end{align}
\end{subequations}
where $k_g \mathcal{P} A^2 = k_g m (m+n-1) P A^2$ and
the prime denotes $\partial_\tau = D^{-1} \partial_t$.
The two constraints are
\begin{subequations}\label{mp1C}
\begin{align}
\label{mp1C1}
C_1 & = (m P + n Q)^2 + k_g m (m+n-1) A^2 - 1 = 0\\
\label{mp1C2}
C_2 & = m P^2 + n Q^2 - k_h n (m+n-1) B^2 -1 = 0\:.
\end{align}
\end{subequations}
We consider the system~\eqref{mp1} on the (new) state space $\bar{\SS}$ defined by 
$C_1 =0$, $C_2 =0$ and $A \geq 0$, $B \geq 0$.
In Eqs.~\eqref{mp1} and~\eqref{mp1C} $k_g =1$ and $k_h =-1$ is understood.
Note that the dynamical system induced on the boundaries $A =0$ and $B= 0$ of the state space
describes the reduced dynamics of the cases $k_g = 0$, $k_h = -1$ and
$k_g =1$, $k_h =0$, respectively.

Equation~\eqref{mp1C2} together with $B \geq 0$ describes the ``northern hemisphere'' of an ellipsoid.
By~\eqref{mp1C1}, $A \geq 0$ implies $-1 \leq (m P + n Q)\leq 1$.
Hence, $\SS$ corresponds to a strip on the ellipsoid that connects two
diametrically opposite segments of the ``equator'' $B=0$ via the ``north pole'' $P = Q =0$.
Topologically, $\SS$ is a rectangle, whose sides are
$B=0$ (with $P >0$), $A=0$ (with $m P +n Q =1$), $B =0$ (with $P <0$), and $A=0$ (with $m P + n Q = -1$);
see Fig.~\ref{fig:mp1}.

\subsection{Equilibrium points and invariant subsets}

The equilibrium points in $\bar{\SS}$ of the autonomous system~\eqref{mp1} are 
\begin{subequations}
\begin{align} 
(\text{F}_{1\ldots 4})&: \quad A = B = 0, \quad m P^2 + n Q^2 = 1, \quad m P+n Q= \pm 1\:, \\
(\text{F}_{\mathrm{A_{+,-}}})&: \quad A = 0, \quad B^2 = \frac{n-1}{n^2} \frac{1}{m+n-1}, \quad P =
0,\quad Q = \pm \frac{1}{n} \, \:.
\end{align} 
\end{subequations}
The equations $m P^2 + n Q^2 = 1$ and $m P+n Q=\pm 1$ possess four solutions $(P_1, Q_1), \ldots, (P_4, Q_4)$,
which define the fixed points $\text{F}_1,\ldots, \text{F}_4$.
For $i=1,2$ we have $m P_i + n Q_i = 1$, 
$P_1 > 0$ ($Q_1 < 0$) and $P_2 < 0$ ($Q_2 > 0$);
for $i=3,4$ we have $m P_i + n Q_i = -1$, see Fig.~\ref{fig:mp1}.
Note that $(P_3, Q_3) = -(P_1,Q_1)$ and $(P_4, Q_4) = -(P_2,Q_2)$.
Clearly, $|P_i| < 1$ and $|Q_i|< 1$, because $m>1$, $n>1$.
We define the boundaries of the state space $\UU_{\mathrm{A}} = \{A=0\} \cap \bar{\SS}$ and
$\UU_{\mathrm{B}} = \{B=0\} \cap \bar{\SS}$.

\subsection{Stability analysis} 

\quad\newline
\noindent\textbf{The case $(\text{F}_{1\ldots 4})$}:
Since the fixed points $(\text{F}_{1\ldots 4})$ are given as intersections of the
one-dimensional invariant subspaces $\UU_{\mathrm{A}}$ and $\UU_{\mathrm{B}}$,
the eigenvectors of the linearization of the system~\eqref{eq:evol}
must be tangential to $\UU_{\mathrm{A}}$ and $\UU_{\mathrm{B}}$.
The relation $A^{-1} A^\prime |_{(\text{F}_{i})} = P_i - \mathrm{sign}(m P_i + n Q_i)$
and the analogous relation for $B$
gives the eigenvalues of the linearization of the dynamical system at the fixed points.
The eigenvalues are $(P_i-1)$ and $(Q_i-1)$ for
$i=1,2$;
hence $(\text{F}_{1, 2})$ are sinks.
The eigenvalues are $(P_i+1)$ and $(Q_i+1)$ for $i=3,4$;
hence $(\text{F}_{3, 4})$ are sources.

\noindent\textbf{The case $(\text{F}_{\mathrm{A_{+,-}}})$}: 
One eigenvector of the linearization of the system at $(\text{F}_{\mathrm{A_{+,-}}})$
is tangential to $\UU_{\mathrm{A}}$.
The associated eigenvalue is $\pm (n-1)/n$, which follows when
we set $A=0$ in~\eqref{mp1P} and compute
$P^{-1} P^\prime|_{(\text{F}_{\mathrm{A}_{+,-}})} = \pm [1 -(1/n)]$. 
The second eigenvector is
transversal to $\UU_{\mathrm{A}}$;
the associated eigenvalue is
$(\mp 1/n)$, since $A^{-1} A^\prime |_{(\text{F}_{\mathrm{A}_{+,-}})} = \mp(1/n)$.
We conclude that $(\text{F}_{\mathrm{A_{+,-}}})$ are saddles, 
so that generic orbits in $\SS$ do not tend to $(\text{F}_{\mathrm{A_{+,-}}})$.
However, there exists exactly one orbit that converges to $(\text{F}_{\mathrm{A}_+})$
along its stable subspace as $\tau\rightarrow \infty$,
and there exists exactly one orbit that converges to $(\text{F}_{\mathrm{A}_-})$
along its unstable subspace as $\tau\rightarrow -\infty$.

\subsection{Global dynamics}

The existence of periodic orbits in $\SS$ is excluded by the absence of interior
fixed points; furthermore, there exist no heteroclinic cycles.
Consequently, all orbits converge to the fixed points as $\tau\rightarrow \pm \infty$.

The fixed points $(\text{F}_{\mathrm{A_{+,-}}})$ are of particular interest. 
Recall that 
there exists exactly one
orbit that originates from $(\text{F}_{\mathrm{A_{-}}})$ into the interior of $\SS$,
and one orbit that converges to $(\text{F}_{\mathrm{A_{+}}})$ as $\tau\rightarrow \infty$.
Making use of global properties of the flow we prove that
these orbits do not coincide:

Consider Eq.~\eqref{mp1P} in the case $|P| \ll |Q|$, $A < B$, i.e.,
\begin{equation}
P^\prime \approx k_g (m+n-1) A^2 + (-k_h) n^2 (m+n-1) B^2 P Q \:.
\end{equation}
In a neighborhood of $(\text{F}_{\mathrm{A_{-}}})$, we obtain
$P^\prime > 0$ if $P\leq 0$.
It follows that the orbit originating from $(\text{F}_{\mathrm{A_{-}}})$
satisfies $P(\tau) > 0$ for all $\tau$ sufficiently small.
Analogously, in a neighborhood of $(\text{F}_{\mathrm{A_{+}}})$,
$P^\prime > 0$ if $P\geq 0$, hence the orbit converging to $(\text{F}_{\mathrm{A_{+}}})$
satisfies $P(\tau) < 0$ for sufficiently large values of $\tau$.
Since $P^\prime > 0$ when $P=0$ by~\eqref{mp1P}, it follows that
the two orbits are distinct.

Using again the fact that $P^\prime >0$ when $P =0$ it is straightforward to
conclude that the orbit that originates from $(\text{F}_{\mathrm{A_{-}}})$
converges to $(\text{F}_1)$ as $\tau\rightarrow \infty$;
conversely there is one orbit that originates from
$(\text{F}_3)$ and ends in $(\text{F}_{\mathrm{A_{+}}})$.

\begin{figure}[Ht]
\begin{center}
\psfrag{P}[cc][cc][1][0]{$P$}
\psfrag{Q}[cc][cc][1][0]{$Q$}
\psfrag{F1}[cc][cc][0.8][0]{$(\text{F}_{1})$}
\psfrag{F2}[cc][cc][0.8][0]{$(\text{F}_{2})$}
\psfrag{F3}[cc][cc][0.8][0]{$(\text{F}_{3})$}
\psfrag{F4}[cc][cc][0.8][0]{$(\text{F}_{4})$}
\psfrag{FA+}[cc][cc][0.8][0]{$(\text{F}_{\mathrm{A_{+}}})$}
\psfrag{FA-}[cc][cc][0.8][0]{$(\text{F}_{\mathrm{A_{-}}})$}
\psfrag{MP-1}[lc][lc][0.7][0]{$m P +n Q=-1$}
\psfrag{MP1}[lc][lc][0.7][0]{$m P + n Q=1$}
\psfrag{B1}[cc][cc][0.8][55]{$B=0$}
\psfrag{B2}[cc][cc][0.8][50]{$B=0$}
\psfrag{A1}[cc][cc][0.8][-30]{$A=0$}
\psfrag{A2}[cc][cc][0.8][-30]{$A=0$}
\includegraphics[width=0.6\textwidth]{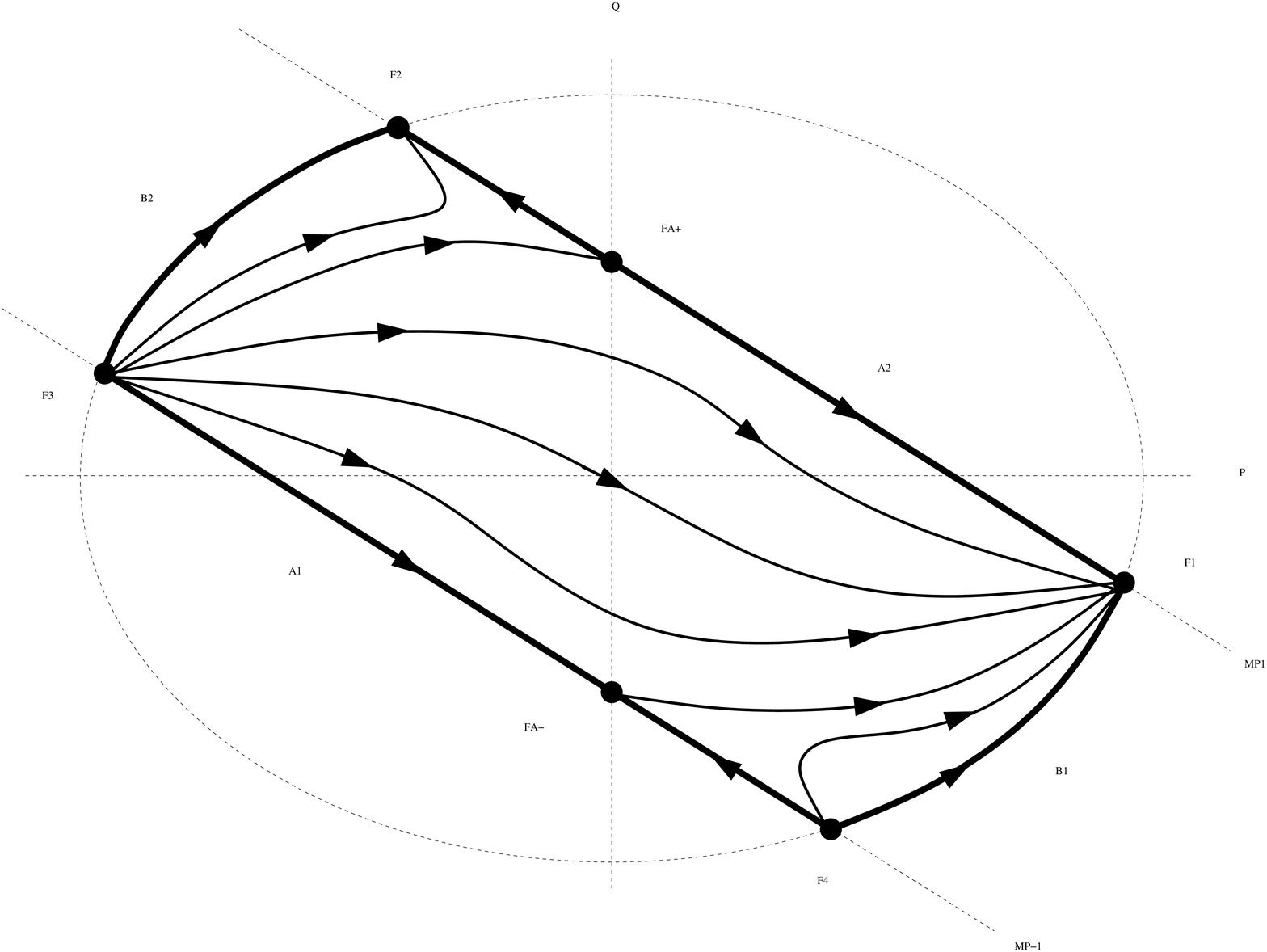}
\caption{A projection of the state space $\SS$ in the case $k_g=1$, $k_h=-1$ and schematic of the (projection of the) flow.}
\label{fig:mp1}
\end{center}
\end{figure}

\subsection{Asymptotics of the scale factors}

\quad\newline
\noindent\textbf{The case $(\text{F}_{1\ldots 4})$}:
Consider a solution that converges to 
the fixed point $(\text{F}_{3})$
as $\tau\rightarrow -\infty$.
In a neighborhood of the fixed point 
the dynamical system~\eqref{mp1} can be approximated by its linearization at $(\text{F}_3)$,
which is given by
\begin{equation}\label{linatF3}
A^\prime = A [P_3 +1]\,, \qquad 
B^\prime = B [Q_3 +1]\,, \qquad
P \equiv P_3\,, \qquad 
Q\equiv Q_3 \:.
\end{equation}
The equations~\eqref{linatF3} coincide with the system of equations~\eqref{mp1}~\&~\eqref{mp1C}
obtained for $k_g=k_h =0$.
Consequently,
all solutions that converge to $(\text{F}_3)$ as $\tau\rightarrow -\infty$
behave asymptotically like
solutions representing a spacetime where both factors $g$ and $h$ are Ricci flat.
In other words, the dynamical
effects of curvature become negligible in the limit.

Solving~\eqref{linatF3} and the decoupled equation~\eqref{mp1D}
we obtain
\begin{equation}
a = \bar{a}_0 \: t^{P_1} \,,\qquad b = \bar{b}_0\: t^{Q_1} 
\end{equation}
for $t\searrow 0$; $\bar{a}_0$ and $\bar{b}_0$ are constants.
We have used here that $-P_3 = P_1$ and $-Q_3 = Q_1$;
recall that $m P_1 + n Q_1 =1$, $m P_1^2 + n Q_1^2 =1$, $P_1 >0$.
A completely analogous result is obtained for solutions that approach
$(\text{F}_4)$, where $(P_1,Q_1)$ is replaced by $(P_2,Q_2)$.

Similarly, solutions that tend to $(\text{F}_{1})$ as $\tau\rightarrow \infty$ 
correspond to 
\begin{equation}\label{mp1aforF1}
a = a_0 \:(T-t)^{P_1} \,,\qquad  b = b_0 \:(T-t)^{Q_1}\:;
\end{equation}
the analogous result holds for $(\text{F}_{2})$.
In~\eqref{mp1aforF1}, $T$ is the time of the future singularity of the spacetime.

\noindent\textbf{The case $(\text{F}_{\mathrm{A_{+,-}}})$}: 
Now consider a solution along the orbit 
that converges to $(\text{F}_{\mathrm{A}_+})$ as $\tau\rightarrow \infty$.
We find
\begin{equation}
a \rightarrow a_0 \,\qquad b = \frac{\sqrt{m+n-1}}{\sqrt{n-1}}\, (T-t) 
\end{equation}
as $t\rightarrow T$.
Analogously,
\begin{equation}
a \rightarrow \bar{a}_0 \,\qquad b = \frac{\sqrt{m+n-1}}{\sqrt{n-1}}\, t 
\end{equation}
for a solution along the orbit that converges to $(\text{F}_{\mathrm{A}_-})$ as $\tau\rightarrow -\infty$.

The global properties of the flow of the dynamical system translate to the following statements:
all solutions undergo recollapse; there exist two generic types of past singularities 
and one special type; analogously, there exist two generic types of future singularities and one special type, 
cf.~Fig.~\ref{fig:mp1}.
(I) Every solution with a past singularity of the type ($a\propto t^{P_2}$, $b\propto t^{Q_2}$)
possesses a future singularity of the
type ($a\propto (T-t)^{P_1}$, $b\propto (T-t)^{Q_1}$). 
(II) For a solution with a past singularity of the type
($a\propto t^{P_1}$, $b\propto t^{Q_1}$) there exist three scenarios:
(1) ($a\propto (T-t)^{P_1}$, $b\propto (T-t)^{Q_1}$),
(2) ($a\propto (T-t)^{P_2}$, $b\propto (T-t)^{Q_2}$), or 
(3) ($a \rightarrow \mathrm{const}$, $b \propto (T-t)$) as $t\rightarrow T$.
The scenario (3) is not generic.
(III) There exists a non-generic solution with a past singularity of the type
($a\rightarrow \mathrm{const}$, $b\propto t$) for $t\rightarrow 0$.
This solution possesses a future singularity of the type 
($a\propto (T-t)^{P_1}$, $b\propto (T-t)^{Q_1}$).

\subsection{Accelerating cosmologies from compactification} 

\quad\newline
Consider, in complete analogy to Section~\ref{acceleration}, the dimensionally reduced spacetime
$\mathbb{R} \times M$ with metric
\begin{equation}
\gamma = -d\bar{t}^2 + \bar{a}^2\, g \:,
\end{equation}
where $d\bar{t}/d t = b^{n/(m-1)}$ and $\bar{a} = a b^{n/(m-1)}$. 
Orbits in $\SS$ generate cosmological models $(\mathbb{R}\times M, \gamma)$; in the following
we analyze their properties.

\begin{figure}[Ht]
\begin{center}
\psfrag{P}[cc][cc][1][0]{$P$}
\psfrag{Q}[cc][cc][1][0]{$Q$}
\psfrag{F1}[cc][cc][0.8][0]{$(\text{F}_{1})$}
\psfrag{F2}[cc][cc][0.8][0]{$(\text{F}_{2})$}
\psfrag{F3}[cc][cc][0.8][0]{$(\text{F}_{3})$}
\psfrag{F4}[cc][cc][0.8][0]{$(\text{F}_{4})$}
\psfrag{FA+}[cc][cc][0.8][0]{$(\text{F}_{\mathrm{A_{+}}})$}
\psfrag{FA-}[cc][cc][0.8][0]{$(\text{F}_{\mathrm{A_{-}}})$}
\psfrag{MP-1}[lc][lc][0.7][0]{$m P +n Q=-1$}
\psfrag{MP1}[lc][lc][0.7][0]{$m P + n Q=1$}
\psfrag{B1}[cc][cc][0.8][55]{$B=0$}
\psfrag{B2}[cc][cc][0.8][50]{$B=0$}
\psfrag{A1}[cc][cc][0.8][-30]{$A=0$}
\psfrag{A2}[cc][cc][0.8][-30]{$A=0$}
\psfrag{A}[cc][cc][1.5][0]{$\AA_{\mathrm{E}}$}
\psfrag{B}[cc][cc][1.5][0]{$\AA_{\mathrm{C}}$}
\includegraphics[width=0.6\textwidth]{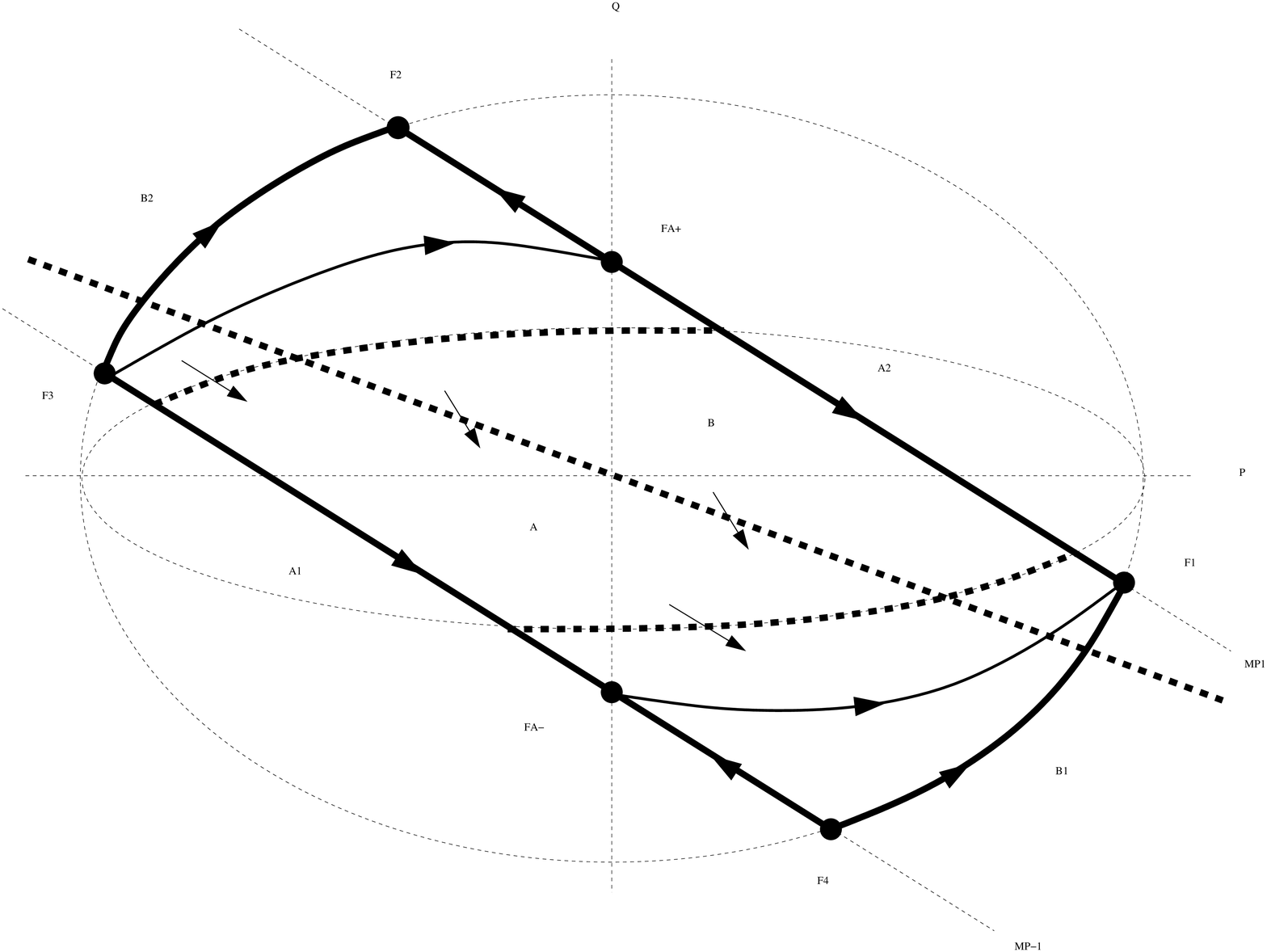}
\caption{The state space $\SS$ in the case $k_g=1$, $k_h=-1$.}
\label{fig:mp1accel}
\end{center}
\end{figure}

A simple calculation shows that the derivative of the scale factor $\bar{a}$ is 
\begin{equation}\label{baraonemp1}
\frac{d \bar{a}}{d \bar{t}} = -\frac{1}{m-1} \, \frac{1}{A}\: [ (m-1) P + n Q]\:;
\end{equation}
hence the equation $[(m-1) P + n Q]=0$ divides the state space:
there exists a domain of expansion and a domain of contraction in $\SS$;
the former is the region below the straight dashed line in Fig.~\ref{fig:mp1accel}.
As long as an orbit lies in this domain, the cosmological model it represents
is expanding.
Superimposing Figs.~\ref{fig:mp1} and~\ref{fig:mp1accel} we see that
all models $(\mathbb{R}\times M,\gamma)$ are expanding initially and contracting toward the end.

Differentiating~\eqref{baraonemp1} we obtain
\begin{equation}\label{baratwomp1}
\frac{d^2 \bar{a}}{d\bar{t}^2} = \frac{a b^{-n/(m-1)}}{m-1} D^2  
\Big[\frac{1}{m} -P^2 - \frac{n}{m} (m+n) Q^2 \Big]\:.
\end{equation}
It is straightforward to show that $d^2 \bar{a}/d\bar{t}^2$ is positive in some
domain $\AA$ of the state space $\SS$. This domain is indicated by curved dashed lines in
Fig.~\ref{fig:mp1accel}.
The intersection with the domain of expansion is non-empty, thus we 
see that there exists 
a domain $\AA_{\mathrm{E}} \subseteq \SS$ of accelerated expansion:
whenever a solution passes through this domain, the cosmological model $(\mathbb{R}\times M, \gamma)$ it represents
undergoes accelerated expansion. (In the complementary domain $\AA_{\mathrm{C}}$ we have
$d\bar{a}/d\bar{t} < 0$ and $d^2 \bar{a}/d \bar{t}^2 > 0$.)

The boundary of $\AA_{\mathrm{E}}$ consists of semipermeable segments. This is because
\begin{equation}
\frac{\partial}{\partial \tau} \left(\frac{d^2 \bar{a}}{d\bar{t}^2} \right) 
\Big|_{d^2 \bar{a}/d\bar{t}^2 =0} 
\propto -(1- m P^2 -n Q^2) (P-Q) 
\end{equation}
and
\begin{equation}
\frac{\partial}{\partial \tau} \left(\frac{d \bar{a}}{d\bar{t}} \right) 
\Big|_{d \bar{a}/d\bar{t} =0}
\propto (m-1) A^2 - n B^2\:.
\end{equation}
Hence, orbits can enter $\AA_{\mathrm{E}}$ through the upper or the right boundary
and leave through the lower boundary.

We conclude that there exists a one-parameter family of solutions that  
exhibits a phase of accelerated expansion, cf.~Figs.~\ref{fig:mp1} and~\ref{fig:mp1accel}; 
these solutions are a subfamily of
the family of solutions that connects $(\text{F}_{3})$ with $(\text{F}_{1})$.
Note that there can be at most one phase of
accelerated expansion.
Interestingly enough,
not all solutions are simply expanding-contracting; there exist solutions 
with two phases of expansion, i.e., models that are expanding-contracting-expanding-contracting.
For those models the 
second 
expanding phase begins as a phase of accelerated expansion.

\section{Concluding remarks}

In this paper we have 
analyzed the dynamics of $D$-dimensional vacuum doubly warped product
spacetimes and their dimensional reduction to the conformal Einstein frame.
The method of scale invariant dynamics has allowed us to
give a comprehensive description of the global dynamics; in particular,
we have proved that, for $D \geq 10$, when the $4$-dimensional
spacetime is a $\kappa=-1$ Friedmann model and the internal space a negatively curved
Einstein space, there exists a unique (thus non-generic)
dimensionally reduced model which is expanding at an acclerated rate
for all times (eternal acceleration). 
Generic models, on the other hand, fall into classes:  
there exist models that exhibit a transient phase of acceleration, late time acceleration,
or no accelerated expansion at all.

The $D$-dimensional models we have considered are of the simplest type:
doubly warped product solutions of the Einstein vacuum equations.
Generalizations include multiply warped product spacetimes, see, e.g.,
\cite{Chen/Ho/Neupane/Ohta/Wang:2003}, $D$-dimensional gravity coupled
to $n$-form field strengths~\cite{Jarv/Mohaupt/Saueressig:2004} or
$n$-form field strengths and a dilaton, see, e.g., \cite{Ohta:2003, Roy:2003};
for reviews see~\cite{Wohlfarth:2004, Ohta:2005}.

In the study of these models, in particular in view of
the question of accelerated expansion, 
dynamical systems methods and phase space analysis have proved 
to be powerful tools;
for instance, the methods permit
the complete classification of the qualitative late-time
bahavior of models~\cite{Wohlfarth:2004, Jarv/Mohaupt/Saueressig:2004}
(as re-derived by the local dynamical systems analysis
in Section~\ref{sec:solutions} and at the end of Section~\ref{acceleration}).
In the present article, in order to obtain a description of the global dynamics of models 
we have employed 
the method of scale-invariant dynamics.
This approach naturally embeds the local analysis 
(late time behavior, \ldots) into a global context, which
has allowed us, in particular, to establish the existence of
a non-generic model exhibiting eternal acceleration.
Whether the methods generalize to 
models as in \cite{Chen/Ho/Neupane/Ohta/Wang:2003, Ohta:2003, Jarv/Mohaupt/Saueressig:2004}
is under current investigation.

\subsection*{Acknowledgements} We thank Vince Moncrief, Alan Rendall
and Mattias Wohlfarth for useful discussions and Paul Tod for pointing us to
references~\cite{Dancer/Wang:1999,Dancer/Wang:2001}.
The authors are grateful for the 
hospitality and support of the Isaac Newton Institute in Cambridge, where part of the work on
this paper was performed.

\vspace{3cm}



\end{document}